\documentclass[a4paper,twocolumn,10pt,accepted=2023-01-10]{quantumarticle}
\pdfoutput=1
\usepackage[utf8]{inputenc} 
\usepackage[english]{babel} 
\usepackage[T1]{fontenc}
\usepackage{hyperref}
\usepackage{graphicx}  
\usepackage[numbers,sort&compress]{natbib}
\usepackage{subfigure}
\usepackage{amssymb}   
\usepackage{amsmath}
\usepackage{mathtools}
\usepackage{color}
\usepackage[normalem]{ulem}
\usepackage{multirow}
\usepackage{tabulary}
\newcolumntype{K}[1]{>{\centering\arraybackslash}p{#1}}

\newcommand{\bra}[1]{\langle #1|}
\newcommand{\ket}[1]{| #1 \rangle}

\begin{document}

\title{Symmetry enhanced variational quantum spin eigensolver}

\author{Chufan Lyu}
\affiliation{Institute of Fundamental and Frontier Sciences, University of Electronic Science and Technology of China, Chengdu 610051, China}

\author{Xusheng Xu}
\affiliation{Central Research Institute, 2012 Labs, Huawei Technologies}

\author{Man-Hong Yung}
\affiliation{Central Research Institute, 2012 Labs, Huawei Technologies}
\affiliation{Department of Physics, Southern University of Science and Technology, Shenzhen 518055, China}
\affiliation{Shenzhen Institute for Quantum Science and Engineering, Southern University of Science and Technology, Shenzhen 518055, China}

\author{Abolfazl Bayat}
\email{abolfazl.bayat@uestc.edu.cn}
\affiliation{Institute of Fundamental and Frontier Sciences, University of Electronic Science and Technology of China, Chengdu 610051, China}

\begin{abstract}
The variational quantum-classical algorithms are the most promising approach for achieving quantum advantage on near-term quantum simulators. Among these methods, the variational quantum eigensolver has attracted a lot of attention in recent years. While it is very effective for simulating the ground state of many-body systems, its generalization to excited states becomes very resource demanding. Here, we show that this issue can significantly be improved by exploiting the symmetries of the Hamiltonian. The improvement is even more effective for higher energy eigenstates. We introduce two methods for incorporating the symmetries. In the first approach, called hardware symmetry preserving, all the symmetries are included in the design of the circuit. In the second approach, the cost function is updated to include the symmetries. The hardware symmetry preserving approach indeed outperforms the second approach. However, integrating all symmetries in the design of the circuit could be extremely challenging. Therefore, we introduce hybrid symmetry preserving method in which symmetries are divided between the circuit and the classical cost function. This allows to harness the advantage of symmetries while preventing sophisticated circuit design. 

\end{abstract}

\maketitle

\section{Introduction}\label{section:Introduction}

The computation power provided by the emerging quantum simulators and computers will fundamentally transform our technology across different disciplines, including condensed matter physics~\cite{kokail2019self}, chemistry~\cite{aspuru2005simulated}, material design~\cite{helgaker2014molecular} and finance~\cite{orus2019quantum,rebentrost2018quantum,egger2020quantum}. 
Thanks to recent advancements in quantum technologies, quantum simulators are rapidly emerging in various physical platforms~\cite{bordia2017probing,schreiber2015observation,gross2017quantum,hempel2018quantum,lanyon2011universal,aspuru2012photonic,wang2020integrated,hensgens2017quantum,salfi2016quantum,arute2020hartree,barends2016digitized}. However, the current Noisy Intermediate Scale Quantum (NISQ) simulators suffer from imperfect initialization, noisy operations and faulty readout~\cite{preskill2018quantum}. 
Thus, developing novel algorithms which are friendly to imperfect NISQ simulators and capable of achieving quantum advantage has attracted a lot of attention in recent years~\cite{bharti2021noisy}. An important class of these algorithms are variational methods which are performed on a hybrid of NISQ simulators and classical optimizers~\cite{peruzzo2014variational,cerezo2021variational,mcclean2016theory,yuan2019theory,xin2020quantum}. In such algorithms, a cost function is measured on a parameterized quantum circuit. Then the result is fed into a classical optimizer to update the parameters of the circuit. The loop is repeated until the cost function is minimized. Therefore, all variational quantum-classical algorithms demand two types of resources: (i) quantum resources which is quantified through either the circuit depth or equivalently the number of gates; and (ii) classical resources which is quantified through the convergence speed. So far, these variational algorithms have been developed for addressing problems in quantum machine learning~\cite{biamonte2017quantum,arunachalam2017survey,ciliberto2018quantum,dunjko2018machine,farhi2018classification,schuld2019quantum}, combinatorial optimization~\cite{farhi2014quantum,bravyi2020obstacles}, dynamical simulations in closed~\cite{cirstoiu2020variational,gibbs2021longtime,yuan2019theory,mcardle2019variational,heya2019subspace} and open~\cite{huh2014linear,hu2020quantum,endo2020variational,yuan2019theory,haug2020generalized} systems, quantum sensing~\cite{meyer2021variational,meyer2021fisher,beckey2020variational,kaubruegger2019variational,koczor2020variational,ma2021adaptive,haug2021natural}, computational chemistry~\cite{cao2021larger,arute2020hartree,peruzzo2014variational,kandala2017hardware,nam2020ground} and condensed matter physics~\cite{BravoPrieto2020scalingof,Lyu2020accelerated,uvarov2020variational,okada2022identification}. 

The most popular application of variational quantum algorithms is to find the ground state of complex many-body systems. For fulfilling this task, the Variational Quantum Eigensolver (VQE) algorithm has been designed to target the ground state of a many-body system through minimizing the average energy~\cite{peruzzo2014variational,mcclean2016theory}. The VQE has been extensively applied to quantum chemistry problems~\cite{arute2020hartree,cao2021larger,kandala2017hardware,nam2020ground} and experimentally realized on superconducting~\cite{arute2020hartree,kandala2017hardware,chen2020demonstration,harrigan2021quantum} and ion trap~\cite{kokail2019self,hempel2018quantum,pagano2020quantum} quantum simulators. Several attempts have been made to enhance the VQE performance, including: minimizing the number of required measurements~\cite{zhao2020measurement,izmaylov2019unitary,verteletskyi2020measurement,gokhale2020n,ralli2021implementation,van2021measurement}, improving the initialization~\cite{Grant2019initialization,volkoff2021large,Lyu2020accelerated}, speeding up the classical optimization~\cite{stokes2020quantum,khairy2020learning,gilyen2019optimizing} and designing better circuits~\cite{ostaszewski2021reinforcement,pirhooshyaran2020quantum,fosel2021quantum,rattew2019domain,chivilikhin2020mogvqe,huang2022robust}. 
Several important phenomena in physics, such as topological phases~\cite{asboth2016schrieffer}, are  described by the knowledge of a few low-energy eigenstates and not just the ground state. Therefore,
the generalization of VQE to higher energy eigenstates is of high importance. However, the existing VQE protocols for simulating excited states~\cite{nakanishi2019subspace,higgott2019variational,mcclean2017hybrid,santagati2018witnessing} are very resource demanding which makes their scalability and practicality in doubt.  
Therefore, improving the performance of VQE in terms of feasibility for simulating large systems is a key milestone to achieve quantum advantage with NISQ devices. 

Symmetry is one of the most profound concepts in physics, especially in quantum mechanics~\cite{greiner2012quantum}. Most physical systems reveal various forms of symmetries for which a precise mathematical description has been developed~\cite{mcweeny2002symmetry}. 
Symmetries have been exploited to improve data mitigation~\cite{sagastizabal2019experimental}, quantum machine learning~\cite{Meyer2022exploiting}, quantum state representation~\cite{liu2019variational} and variational quantum optimization~\cite{bravyi2020obstacles} in NISQ devices. In addition, the VQE algorithm can also hugely benefit from the incorporation of symmetries. There are two ways to incorporate symmetries in the VQE algorithms for simulating the ground state: (i) designing the circuit to naturally generate the quantum states with the relevant symmetry~\cite{barkoutsos2018quantum,wang2009efficient,Lyu2020accelerated,seki2020symmetry,Gard2020,barron2021preserving,zhang2021shallow,zheng2021speeding}; and (ii) adding extra terms to the cost function to penalize the quantum states without the relevant symmetry~\cite{mcclean2016theory,ryabinkin2018constrained}. Two key open problems still exist. Firstly, whether the symmetries can also be exploited for efficiently simulating the excited states. Secondly, which of the above methods, or a hybrid combination of them, are more effective for incorporating symmetries in the VQE algorithm for enhancing its performance.

In this paper, we show that symmetries can indeed significantly improve the VQE for simultaneously simulating several low-energy eigenstates. The improvement becomes even more pronounced for excited states. We introduce two different approaches for incorporating symmetries. First, in hardware symmetry preserving method, we include all the symmetries in the circuit. Second, we add symmetries as proper penalization terms to the cost function. Interestingly, our analysis shows that the first method is more effective with respect to both quantum and classical resources. However, designing a circuit which can integrate all the symmetries can be notoriously difficult. Hence, we introduce hybrid symmetry preserving method in which the two approaches are combined in order to harness the symmetries while keeping the circuit simple. Thanks to significant enhancement in resource efficiency, our proposal paves the way for achieving quantum advantage. In addition, it is very timely and can be implemented on existing quantum simulators.

\section{Ground state VQE}

In this section we provide a brief review on the VQE algorithm for preparing the ground state of a given Hamiltonian using a shallow quantum circuit~\cite{peruzzo2014variational}. In the VQE algorithm, a parameterized quantum circuit, represented by  unitary operator $U(\vec{\theta})$, is used to prepare a quantum state $|\psi(\vec{\theta})\rangle=U(\vec{\theta})\ket{\psi_0}$ for a given $N$ qubits. This parameterized quantum circuit is often referred to as the ansatz, with $\vec{\theta}=(\theta_1, \theta_2, \dots, \theta_L)$ being some tunable parameters in the circuit and $\ket{\psi_0}$ is the input state. By varying $\vec{\theta}$ one can explore some part of the Hilbert space, which is called reachable set. In very deep circuits, and thus large number of parameters $L$, one may generate any possible quantum state of $N$ qubits and thus the reachable set will be the entire Hilbert space. However, we would like to keep the circuit as shallow as possible and restrict ourselves to the most relevant part of the Hilbert space. In particular, in VQE algorithms we are interested in a fairly shallow ansatz for which the reachable set contains the ground state of the Hamiltonian of interest $H$. So far, several choices of the ansatz with different levels of complexity have been proposed~\cite{kandala2017hardware,taube2006new,o2016scalable,romero2018strategies,farhi2014quantum,wecker2015progress,mcclean2016theory}. After choosing the ansatz, one can measure the average energy $\langle H \rangle = \langle \psi(\vec{\theta}) | H | \psi(\vec{\theta}) \rangle$ through some appropriate measurements on the quantum device. This measured average energy will then be fed into a classical optimizer to be minimized through adaptively updating the parameters $\vec{\theta}$ in the quantum circuit. Eventually, the optimization will be finished by obtaining the optimal parameters $\vec{\theta}^{*}$. If the reachable set contains the ground state then the output of the circuit $|\psi(\vec{\theta}^{*})\rangle$ will be the ground state of $H$. 

The conjecture behind the VQE is that a shallow circuit is enough to realize the ground state of the Hamiltonian. The price for keeping the circuit shallow, i.e. saving quantum resources, is to add a classical optimizer which demands extra classical resources. In this paper, we use L-BFGS algorithm as the classical optimizer~\cite{Liu1989}. If the optimization landscape suffers from Barren plateau~\cite{mcclean2018barren,nakata2017unitary}  or the presence of significant number of local minima then the classical optimization may converge very slowly or even never reach the right quantum state. Therefore, in any VQE algorithm it is crucial to quantify both quantum and classical resources. Since single qubit operations are almost perfect, we use the number of two-qubit entangling gates (e.g. controlled-not) in our circuit as a quantification of quantum resources. On the other hand, for classical resources one has to notice that $L$ parameters have to be optimized iteratively. Therefore, a logical definition for Classical Resources ($C_R$) can be the multiplication of the number of parameters $L$ and the number of optimization iterations $n_I$, namely \begin{equation}
    C_R=L \times n_I 
    \label{CR_eq}
\end{equation}

\begin{figure*}[ht]
  \centering

  \includegraphics[width=1.6\columnwidth]{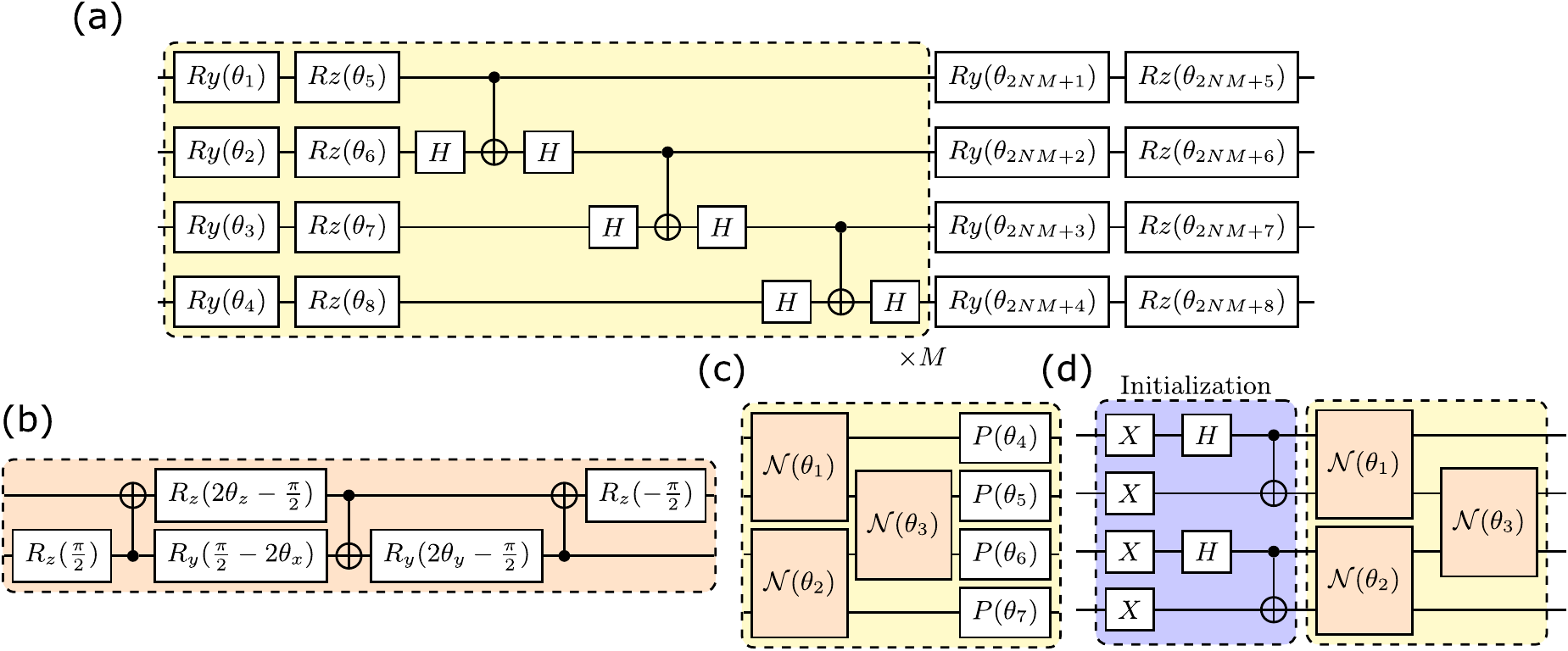}
  \caption{(a) The hardware-efficient ansatz circuit for system size $N=4$. By repeating the circuits shown in yellow, the expressibility of the variational circuit increases. (b) Circuit for realizing $\mathcal{N}(\theta_x, \theta_y, \theta_z)$ as the entangling gate between two qubits. (c) A single layer circuit of $S_z$-conserving ansatz for Heisenberg Hamiltonian of system size $N=4$. (d) A single layer circuit of $S_{tot}$-conserving ansatz is shown in the yellow box. The initialization circuit, shown in blue box, varies for every choice of total spin $s$. Here, the circuit represents the initial state $\ket{\psi_0}=\otimes^{N/2} \ket{\psi^-}$ for $s=0$.}
  \label{fig:ansatz}
\end{figure*}

The choice of ansatz is very crucial in all variational quantum algorithms. Perhaps the most widely used quantum circuit in the literature is the hardware efficient circuit~\cite{kandala2017hardware} which is schematically shown in Fig.~\ref{fig:ansatz}(a). In this circuit the single qubit rotations are defined as
\begin{equation}
    R_{\alpha}(\theta)=e^{i\theta \sigma_\alpha}
\end{equation}
where $\sigma_{\alpha}$ is the Pauli operator with $\alpha{=}x,y,z$. In this paper, we use  Controlled-Not (CNOT) gate as the two-qubit entangling gate in our circuits which is defined as 
\begin{equation}
    CNOT=\begin{pmatrix}
        1 & 0 & 0 & 0 \\
        0 & 1 & 0 & 0 \\
        0 & 0 & 0 & 1 \\
        0 & 0 & 1 & 0 
\end{pmatrix}
\end{equation}

Although, the hardware efficient circuit has been heavily used for solving various problems, it suffers from Barren plateau~\cite{mcclean2018barren,nakata2017unitary} which makes its training extremely difficult, in particular, when the number of layers increases. In addition, hardware efficient circuit does not conserve any symmetry. Therefore, as an alternative, one can use a more complex entangling gate with tunable parameters such as~\cite{Vatan2004}
\begin{equation}
    \mathcal{N}(\theta_x,\theta_y,\theta_z)=e^{i\left(\theta_x \sigma_x^1\sigma_x^2+ \theta_y \sigma_y^1\sigma_y^2+\theta_z \sigma_z^1\sigma_z^2\right)}
\end{equation}
where $\sigma_{x,y,z}^j$ are the Pauli matrices acting on qubit $j$. The circuit of this unitary operation is depicted in 
Fig.~\ref{fig:ansatz}(b). For the special case of $\theta_x{=}\theta_y{=}\theta_z$ this unitary operator conserves the number of excitations as well as the total spin. By combining these two qubit gates and phase shift gates $P(\theta)=\begin{pmatrix}
        1 & 0 \\
        0 & e^{i\theta}
\end{pmatrix}$, as shown in Fig.~\ref{fig:ansatz}(c), one can make a quantum circuit $U(\vec{\theta})$ which conserves  the number of excitation, namely $[U(\vec{\theta}), S_z]=0$, where $S_\alpha=\frac{1}{2}\sum_i \sigma_{\alpha}^i (\alpha=x,y,z)$. In the absence of phase shift gates, see Fig.~\ref{fig:ansatz}(d), this quantum circuit can also preserve the total spin, namely $[U(\vec{\theta}), S_{tot}^2]=0$, where $S_{tot}^2 = S_x^2 + S_y^2 + S_z^2$.

Note that, the convergence of the VQE would be successful only when the energy gap between the ground state and first excited state is larger than the standard deviation of the Hamiltonian with respect to state of the simulator, namely 
\begin{equation}
  \sqrt{\bra{\psi(\vec{\theta}^*)} H^2 \ket{\psi(\vec{\theta}^*)} - {\bra{\psi(\vec{\theta}^*)} H \ket{\psi(\vec{\theta}^*)}}^2 } \leq \Delta E
\end{equation}
where $\Delta E = E_2 - E_1$ is the energy gap between the ground state (with energy $E_1$) and the first excited state (with energy $E_2$). The situation becomes even more serious when there are degenerate eigenstates. In these cases, VQE algorithm may converge to an arbitrary superposition of the degenerate eigenvectors. The superposition may vary from one VQE implementation to another depending on the choice of the initial state or the circuit parameters. However, degeneracy among the eigenvectors is usually a consequence of the presence of some symmetries in the system. Indeed, degenerate eigenstates, despite having the same energy, have different eigenvalues with respect to the symmetry operators. Despite some attempts for exploiting the symmetry properties either through updating the cost function~\cite{mcclean2016theory}, or incorporating the symmetries into the design of the circuit~\cite{Gard2020}, a systematic analysis in this issue is still missing.  
In addition, a few existing proposals~\cite{higgott2019variational,nakanishi2019subspace}, its generalization for higher energy excited states is very challenging.  In the following sections, we try to address these two major issues through exploiting the inherent symmetries in the target Hamiltonian.

\section{Excited state VQE}

The most natural generalization of the VQE is to go beyond the ground state and target higher energy eigenstates. This can be very important in some topological systems in which the topologically relevant states are typically not the ground state~\cite{asboth2016schrieffer}. In addition, in some physical phenomena, e.g. integer and fractional quantum Hall effects, the physics is fully described by low energy spectrum and not just the ground state. 
In order to realize higher energy eigenstates via VQE algorithms, two main methods have been developed: (i) penalty approach~\cite{higgott2019variational}; (ii) subspace-search VQE~(SSVQE) algorithm~\cite{nakanishi2019subspace}. In this section we briefly review these two methods and discuss their pros and cons. 

In the penalty method, we assume that the first $k-1$ lowest energy eigenstates are known through different VQE circuits. Then one can generate the $k$-th eigenstate of the Hamiltonian by penalizing the $k-1$ lowest energy eigenstates in the cost function of VQE~\cite{higgott2019variational}. 
In order to target the $k$-th eigenstate of the system one can update the desired Hamiltonian 
\begin{equation}
    H' = H + \sum_{i=1}^{k-1} \beta_{i} \ket{E_i} \bra{E_i}
\end{equation}
where $\beta_{i}$ are some sufficiently large positive scalar for $i$-th eigenstate. By minimizing the cost function $\langle H' \rangle$, VQE yields the $k$-th eigenstate. This method provides a general and systematic excited state preparation algorithm. Nevertheless, the measurement of the projector terms requires calculating the overlaps $|\langle \psi(\vec{\theta}) | E_i \rangle|^2$ (for $i=1,\cdots,k-1$), which are difficult to realize. So far, two methods have been proposed for computing the overlap between two quantum states. The first method requires an extended circuit in which the depth is increased by a factor of $\sim 2$~\cite{havlivcek2019supervised}. The second approach utilizes the swap test~\cite{garcia2013swap,cincio2018learning} which demands doubling the number of qubits and requires complex many-body controlled gates. Considering the limitations in NISQ devices the penalty approach is unlikely to be beneficial in practice. 

The weighted SSVQE provides an alternative method to generate all the $k$ lowest energy eigenstates of a given Hamiltonian $H$~\cite{nakanishi2019subspace}. In this algorithm, one uses a set of $\{|\phi_{i}\rangle\}_{i=1}^{k}$ orthogonal initial states (namely, $\langle \phi_{i} | \phi_{j} \rangle = \delta_{ij}$) 
as the input of a single parameterized quantum circuit, described by the unitary operator $U(\vec{\theta})$. Since the initial states are orthogonal, the outputs $U(\vec{\theta})| \phi_{j} \rangle$, generated by the same circuit, are orthogonal too. In the weighted SSVQE algorithm, one has to minimize the cost function~\cite{nakanishi2019subspace}  
\begin{equation}
    \mathrm{cost} = \sum_{i=1}^{k} w_{i} \langle \phi_{i}| U^{\dagger}(\vec{\theta}) H U(\vec{\theta}) | \phi_{i} \rangle
    \label{ssvqe_cost}
\end{equation}
where $w_1 > w_2 > \cdots > w_k$ are real positive numbers. Minimizing the cost function in Eq.~\eqref{ssvqe_cost}  produces all the $k$ lowest energy eigenstates such that $|E_{i}\rangle = U(\vec{\theta}^{*})|\phi_{i}\rangle$. The major advantage of the weighted SSVQE is that it provides all the $k$ lowest energy eigenstates in one single optimization procedure, without requiring any overlap of quantum states. However, the algorithm becomes more resource demanding as the number of target eigenstates increases.

\section{Hardware symmetry preserving ground state simulation}

\begin{figure}[h]
  \centering
  \includegraphics[width=\columnwidth]{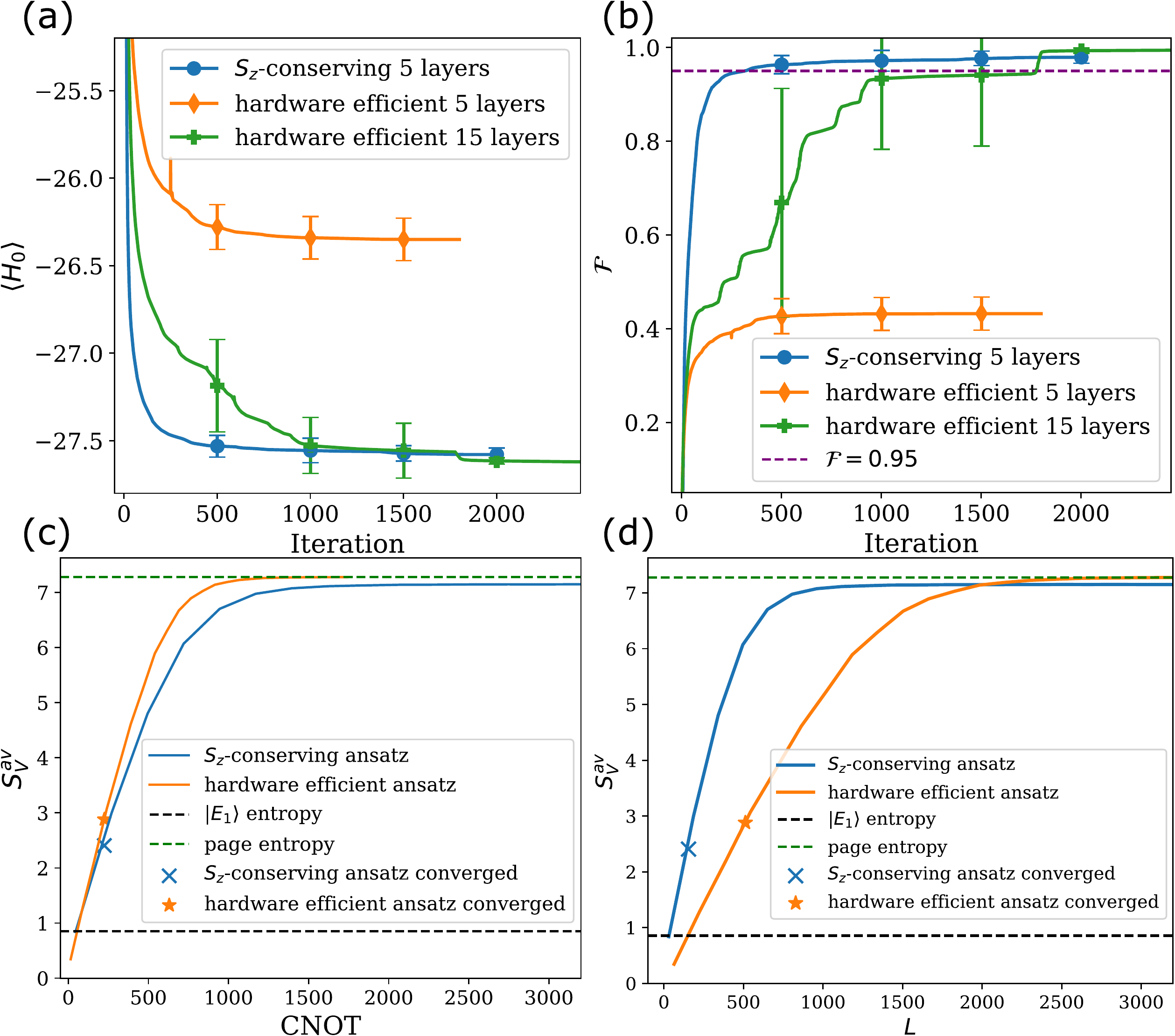}

\caption{The VQE simulation of the ground state of the Heisenberg model with $N=16$ qubits. The average energy, shown in panel (a), and the corresponding obtainable fidelities, shown in panel (b), as a function of optimization iteration $n_I$ for three different circuits. The expressibility  of the circuit, quantified through the average von Neumann entropy $S_V^{av}$, is compared for two different ansatzes versus: (c) the number of CNOT gates, and (d) the number of parameters $L$. The crosses represent the circuit which can achieve fidelity $\mathcal{F} > 0.95$.
}
\label{fig:sz-vs-general_fids}
\end{figure}

In this section, we show how the exploitation of symmetries in the design of the circuit hardware can improve the VQE algorithm for simulating the ground state. We consider a $1$-dimension chain of $N$ spin-$1/2$ particle interacting via  Heisenberg Hamiltonian
\begin{equation}
  H_{0} = J \sum_{i=1}^{N-1} \boldsymbol{\sigma}^{i} \cdot \boldsymbol{\sigma}^{i+1}
  \label{heis_ham}
\end{equation}
where $J=1$ sets the unit of energy and $\boldsymbol{\sigma}^{i} = (\sigma_x^i, \sigma_y^i, \sigma_z^i)$ is the vector of Pauli operators at site $i$. The Heisenberg Hamiltonian in Eq.~\eqref{heis_ham} represents a model of interacting system which cannot be mapped to free fermions and supports several symmetries including the conservation of the total spin, i.e. $[H_0, S_{tot}^2]=0$, as well as its components in all directions, namely $[H_0, S_{\alpha}]=0$.  The first symmetry implies that every eigenstate $\ket{E_k}$ of the system has a specific total spin $s$ which is an integer number for even $N$ or half-integer for odd $N$, such that $\bra{E_k} S_{tot}^2 \ket{E_k}=s(s+1)$. The second symmetry also guarantees that  each eigenstate conserves the $z$ component of the total spin such that $\bra{E_k} S_{z} \ket{E_k}=s_z$, with $-s \le s_z \le s$. In particular, for even $N$, the ground state $\ket{E_1}$ is a global singlet with both $s = 0$ and $s_{z}=0$. Thus we represent it as $\ket{E_{S_1}}=\ket{E_1}$. The first excited state is a global triplet state with total spin $s = 1$ and triply degenerate with $s_z=0,\pm 1$. For the sake of simplicity we use $\ket{E_{T_1}^{(0)}}=\ket{E_2}$, $\ket{E_{T_1}^{(-1)}}=\ket{E_3}$ and $\ket{E_{T_1}^{(+1)}}=\ket{E_4}$. The second excited state is another triplet state with $s=1$ for which we similarly define $\ket{E_{T_2}^{(0)}}=\ket{E_5}$, $\ket{E_{T_2}^{(-1)}}=\ket{E_6}$ and $\ket{E_{T_2}^{(+1)}}=\ket{E_7}$. The fourth eigenstate is a global singlet with $s=0$ which we will represent it as $\ket{E_{S_2}}=\ket{E_8}$.  

We consider the VQE simulation of the ground state of $H_0$ using two different ansatzes, namely: (i) the hardware efficient circuit, shown in Fig.~\ref{fig:ansatz}(a), which conserves no symmetry; and (ii) the $S_z$-conserving circuit, shown in Fig.~\ref{fig:ansatz}(c), which conserves the  number of excitation but do not preserve the total spin. Later, we will also consider circuits which can realize both $S_z$ and $S_{tot}$ symmetries. In both ansatzes, we initialize the circuit in the quantum state $|0,1,0,\cdots,0,1\rangle$. While in the hardware efficient circuit the choice of the initial state does not matter, in the case of $S_z$-conserving circuit this choice is crucial and should have similar $s_z$ as the ground state $\ket{E_{S_1}}$. To see the importance of exploiting the symmetry in the design of the circuit, we perform VQE for a system $N=16$ qubits on both ansatzes.  In Figs.~\ref{fig:sz-vs-general_fids}(a)-(b) we plot the average energy and the obtainable fidelity as a function of optimization iteration, respectively. The error bars are computed for $\sim 100$ repetitions of random samples. Interestingly, while only $5$ layers of excitation conserving circuit is enough for a fast ($\sim 300$ iteration) convergence to $\mathcal{F}>0.95$, the hardware efficient circuit with the same depth can only reach $\mathcal{F} \simeq 0.4$. In fact, the hardware efficient circuit can only reach fidelities above $0.95$ when it contains at least $15$ layers. Even for such a circuit, the optimization needs $\sim 2000$ iterations to converge. This means that the hardware efficient circuit demands way more classical resources ($C_R=870400$ for $\mathcal{F}=0.95$) than the $S_z$-conserving circuit ($C_R=77500$ for $\mathcal{F}=0.95$).

To better understand the difference between the performances of the two circuits one can determine the entangling power of the two circuits in terms of the number of CNOTs, as a quantum resource, as well as the number of parameters $L$, as a quantifier of classical resources (see Eq.~\eqref{CR_eq}). To measure the entangling power, one can compute the average entanglement, quantified through von Neumann entropy, between the two halves of the system. For any quantum state $\ket{\psi(\vec{\theta})}$ at the output of circuit, one can compute the reduced density matrix of the left side of the system by tracing out the $N/2$ qubits on the right side, namely $\rho_{L}(\vec{\theta})=\mathrm{Tr}_R \left[\ket{\psi(\vec{\theta})} \bra{\psi(\vec{\theta})}\right]$. The entanglement between the left and the right side of the system is then quantified by $S_{V}(\vec{\theta})=-\mathrm{Tr}\left[ \rho_{L} \log \rho_L \right]$. The average entangling power of the circuit is then defined as 
\begin{equation}
    S_{V}^{av}=\int d\vec{\theta}~S_{V}(\vec{\theta})
\end{equation}
where the integration is performed over the whole parameter space. For the numerical simulation, we approximate $S_{V}^{av}$ by averaging $S_{V}(\vec{\theta})$ over $500$ random samples of $\vec{\theta}$. In Fig.~\ref{fig:sz-vs-general_fids}(c) we plot $S_{V}^{av}$ versus the number of CNOT gates when the number of layers vary from $1$ to $120$. Note that, for the same number of CNOTs the hardware efficient circuit has many more layers than the $S_z$-conserving circuit. Interestingly, the two ansatzes reach the fidelity $\mathcal{F}=0.95$ when their number of CNOTs are equal to $225$. However, as the figure shows, for this number of CNOTs the hardware efficient have slightly more entangling power than the $S_z$-conserving circuit. Both circuits can reach this fidelity only when their $S_{V}^{av}$ is $\sim 3$ times more than the entanglement in the real ground state. This is due to the fact that both circuits require similar number of CNOTs to reach the same fidelity. It is very insightful to consider the average entangling power $S_{V}^{av}$ versus the number of parameters $L$ as an indicator of required classical resources. In Fig.~\ref{fig:sz-vs-general_fids}(d) we plot $S_{V}^{av}$ as a function of $L$ for the two circuits when the layers vary from $1$ to $120$. Remarkably, the $S_z$-conserving circuit demands much less parameters for reaching the fidelity $\mathcal{F}=0.95$ as it demands a circuit with $L=155$ ($5$ layers) in contrast to $L=512$ ($15$ layers) for the hardware efficient circuit. This clearly shows that for the VQE simulation of the ground state while implementing the symmetry in the hardware may not reduce quantum resources, it significantly enhances classical resource efficiency.

\section{Hybrid symmetry preserving excited state simulation}

\begin{figure}[h]
  \centering
  \includegraphics[width=1\columnwidth]{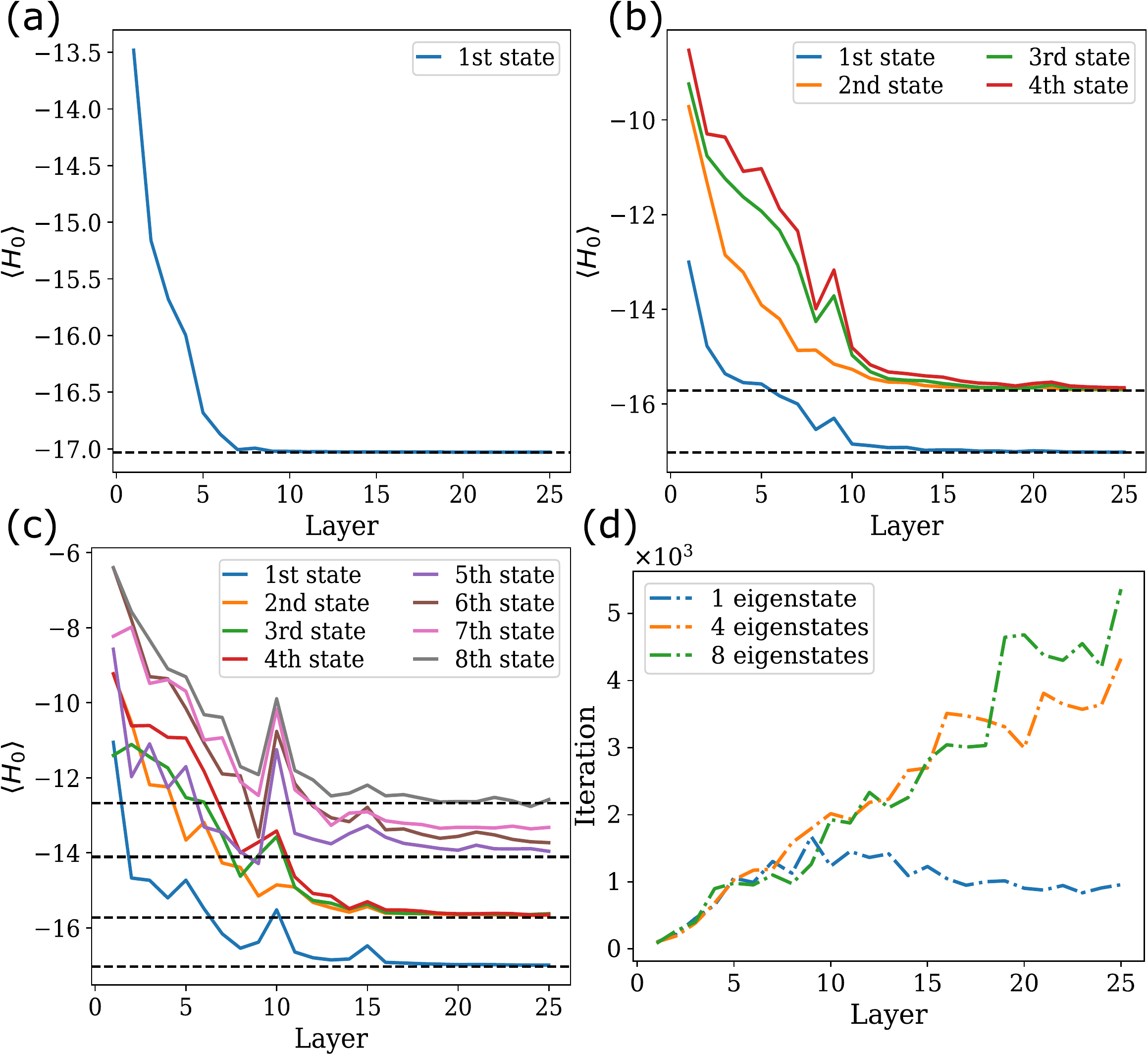}

\caption{The VQE simulation of low-energy eigenstates of the Heisenberg model with system size $N=10$ using hardware efficient ansatz with no symmetry. Different number of eigenstates can simultaneously be targeted in a single circuit using SSVQE. The results for average energy $\langle H_0 \rangle$ as a function of circuit layer are shown for different cases: (a) only the ground state $\ket{E_{S_1}}$ is targeted; (b) four eigenstates including the ground and the three degenerate excited states $\ket{E_{S_1}}$ and $\ket{E_{T_1}^{(0, \pm 1)}}$ are targeted; (c) eight eigenstates including $\ket{E_{S_1}}$, $\ket{E_{T_1}^{(0, \pm 1)}}$, $\ket{E_{T_2}^{(0, \pm 1)}}$ and $\ket{E_{S_2}}$ are targeted. (d) The required optimization iteration corresponding the three previous panels, average over $100$ samples, as a function of circuit layer for various cases in which different number of eigenstates are targeted. 
 }
\label{fig:hard-eff_heis_expt-vs-layer}
\end{figure}

In this section, we focus on the effect of symmetry for the simulation of excited states using the weighted SSVQE. In particular, we consider generating the first $8$ energy eigenstates of the Heisenberg Hamiltonian $H_0$, namely $\ket{E_{S_{1,2}}}$, $\ket{E_{T_{1}}^{(0,\pm 1)}}$ and $\ket{E_{T_{2}}^{(0,\pm 1)}}$. We first use the hardware-efficient ansatz, shown in Fig.~\ref{fig:ansatz}(a), which supports no symmetry. Three different hardware efficient circuits are trained to target $1$ (i.e. $\ket{E_{S_1}}$), $4$ (i.e. $\ket{E_{S_{1}}}$, $\ket{E_{T_{1}}^{(0,\pm 1)}}$) and $8$ (i.e. $\ket{E_{S_{1,2}}}$, $\ket{E_{T_{1}}^{(0,\pm 1)}}$, $\ket{E_{T_{2}}^{(0,\pm 1)}}$) different eigenstates. 
In Figs.~\ref{fig:hard-eff_heis_expt-vs-layer}(a)-(c) we present the average energy $\langle H_0 \rangle$ as a function of circuit layer for targeting $1$, $4$ and $8$ eigenstates, respectively. For each case, the procedure is repeated for $\sim 100$  random initial samples over which the results are averaged to be statistically meaningful. 
As the number of target eigenstates increases, the circuit needs more layers in order to converge for all the corresponding eigenstates. Importantly, as shown in Fig.~\ref{fig:hard-eff_heis_expt-vs-layer}(c), the eigenstates $|E_5\rangle$ to $|E_8\rangle$ do not converge properly even with $25$ layers. In addition to the circuit layer, i.e. quantum resources, the classical optimization also gets more demanding as the number of target eigenstates increases. To quantify the required classical resources, in Fig.~\ref{fig:hard-eff_heis_expt-vs-layer}(d) we plot the number of iterations $n_I$ needed for the convergence of the VQE circuit as a function of circuit layer for various number of target eigenstates.
In order to see how the required resource varies as the number of target eigenstates increases, in Table.~\ref{tbl:minimum_resources} we show the minimum required quantum resources and optimization resources for SSVQE simulation of $k$ eigenstates using hardware efficient ansatz. The number of circuit layers and the number of optimization iterations are chosen such that the energy precision for all the $k$ eigenvectors becomes less than $0.5J$, namely $max\{|\langle H_0 \rangle_i - E_i| \}_{i=1}^k \leq 0.5J$. 
These results clearly show that targeting more eigenstates makes optimization slower and demands more iterations. Thus, one may wonder whether the exploitation of symmetries can help to make the SSVQE more resource efficient.

\begin{table}[h!] 
	\centering
	
\begin{tabular}[c]{ |c|K{0.9cm}|K{0.9cm}|K{1.1cm}|K{1.1cm}|}
	\hline
	& k & $1$ & $4$ & $8$ \\\hline
  \multirow{2}{*}{\parbox{1.9cm}{Quantum Resources}} & Layer & 7 & 18 & 24 \\[1pt]\cline{2-5}
   & CNOT & 63 & 162 & 216 \\ [1pt]\hline
   \multirow{3}{*}{\parbox{1.9cm}{Optimization Resources}} & $n_I$ & 1304 & 3407 & 4205 \\[1pt]\cline{2-5}
   & \multirow{2}{*}{$\mathrm{C_R}$} &  \multirow{2}{*}{208640} &  \multirow{2}{*}{1294660} & \multirow{2}{*}{2102500} \\ [12pt]\hline
\end{tabular}
	\caption{The minimum required quantum resources (namely, the number of layers and the number of CNOT gates), and optimization resources (namely, the number of iterations $n_I$ and the classical resource $\mathrm{C_R}$) for SSVQE simulation of $k$ eigenstates using hardware efficient ansatz. The number of circuit layers and optimization iterations are chosen such that 
  the absolute difference between the average energy and the actual energy becomes less than $0.5J$ (i.e. $max\{|\langle H_0 \rangle_i - E_i| \}_{i=1}^k \leq 0.5J$). }
	\label{tbl:minimum_resources}
\end{table}

Considering one symmetry is not enough to truly distinguish different eigenstates. For instance, $\ket{E_{S_{1,2}}}$ and $\ket{E_{T_{1,2}}^{(0)}}$ all have $s_z=0$. In order to further discriminate the global singlets from the global triplets, one has to also take the total spin $s$ into account. In this section, we use a hybrid approach for implementing the symmetries. Namely, we use the $S_z$-conserving circuit to target quantum states with specific $s_z$ and update the cost function to target quantum states with the right total spin $s$. 

By using the $S_z$-conserving circuit in the SSVQE algorithm one can only target the eigenstates with a given $s_z$. For instance, four eigenstates, namely $|E_{S_{1,2}}\rangle$ and $|E_{T_{1,2}}^{(0)}\rangle$, with $s_z=0$ can be implemented in the same circuit with initial states satisfying $s_z=0$. Similarly, $\ket{E_{T_1}^{(+1)}}$ and $\ket{E_{T_2}^{(+1)}}$ (or equivalently $\ket{E_{T_1}^{(-1)}}$ and $\ket{E_{T_2}^{(-1)}}$) can be targeted on the same $S_z$-conserving circuit with initial states satisfying $s_z=+1$ ($s_z=-1$). This division of eigenstates between different quantum circuits can significantly reduce the circuit complexity in terms of the required number of CNOTs.

\begin{figure}[t]
  \centering
  
  \includegraphics[width=\columnwidth]{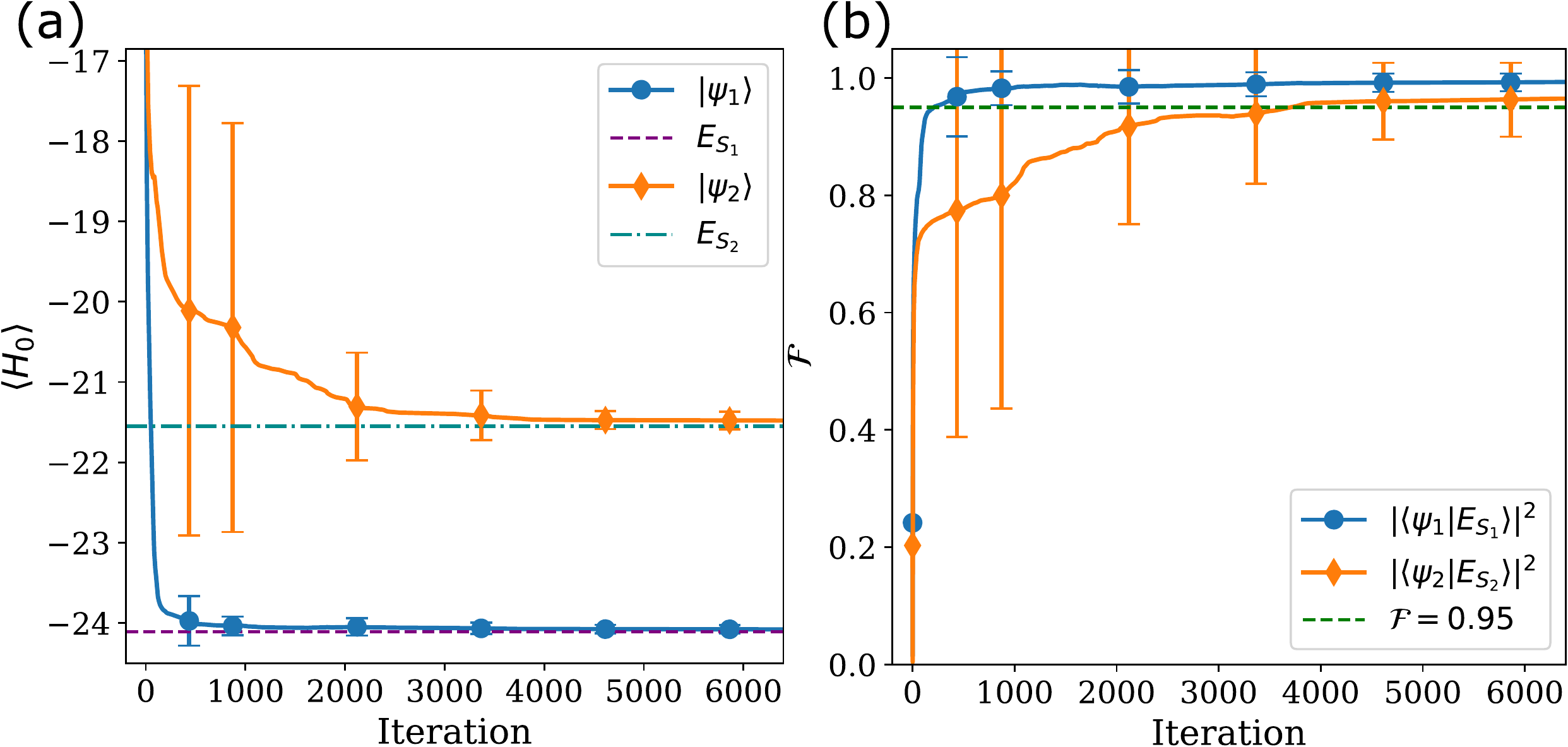}
  
\caption{The SSVQE simulation for simultaneously generating two eigenstates $\ket{E_{S_1}}$ and $\ket{E_{S_2}}$ using $18$ layers of $N=14$ $S_z$-conserving circuit. The average energy, shown in panel (a), and the corresponding obtainable fidelity, shown in panel (b), as a function of optimization iteration $n_I$. The error bars are computed through averaging over $100$ random initial samples.}
\label{fig:heis_e0-e3_result}
\end{figure}

In order to even further simplify the SSVQE circuit one can exploit the total spin symmetry as well. Implementing total spin symmetry in the hardware requires extra CNOT gates, which will be discussed later.  Here, we take a hybrid approach and include the total spin symmetry in the cost function. In particular, for the preparation of $\ket{E_{S_1}}$ and $\ket{E_{S_2}}$ via SSVQE, one can adopt the $S_z$-conserving circuit with two orthogonal initial states $|\phi_1\rangle= |0,1,0,\cdots,1\rangle$ and $|\phi_2\rangle=|1,0,1\cdots,0\rangle$ and the following cost function  
\begin{equation}
    \mathrm{cost} = \sum_{i=1,2} w_i ({\langle H_0 \rangle}_{i} + \beta {\langle S_{tot}^{2} \rangle}_{i}^{2})
    \label{eq:e0+e3_cost}
\end{equation} 
where $\beta=1000$ is a positive constant which is taken to be sufficiently large as in this type of penalty terms the final error in estimation of the cost function is of the order $\mathcal{O}(1/\beta)$~\cite{kuroiwa2021penalty}. 
In Figs.~\ref{fig:heis_e0-e3_result}(a)-(b), we plot the average energy $\langle H_0 \rangle$ and fidelity as a function of optimization iteration, respectively, in a circuit of system size $N=14$ with $18$ layers. The error bars in the figure are computed through averaging over $100$ random initial samples.
Remarkably, both eigenstates can be generated with fidelity above $0.95$ after only $\sim 500$ iterations. This is much better performance than the hardware efficient circuit, see Fig.~\ref{fig:hard-eff_heis_expt-vs-layer}(c), in which  the output did not even converge properly to $\ket{E_{S_2}}$ for a smaller system of size $N=10$ with even $25$ layers and $\sim 5000$ iterations.

\begin{figure}[t]
  \centering
  \includegraphics[width=\columnwidth]{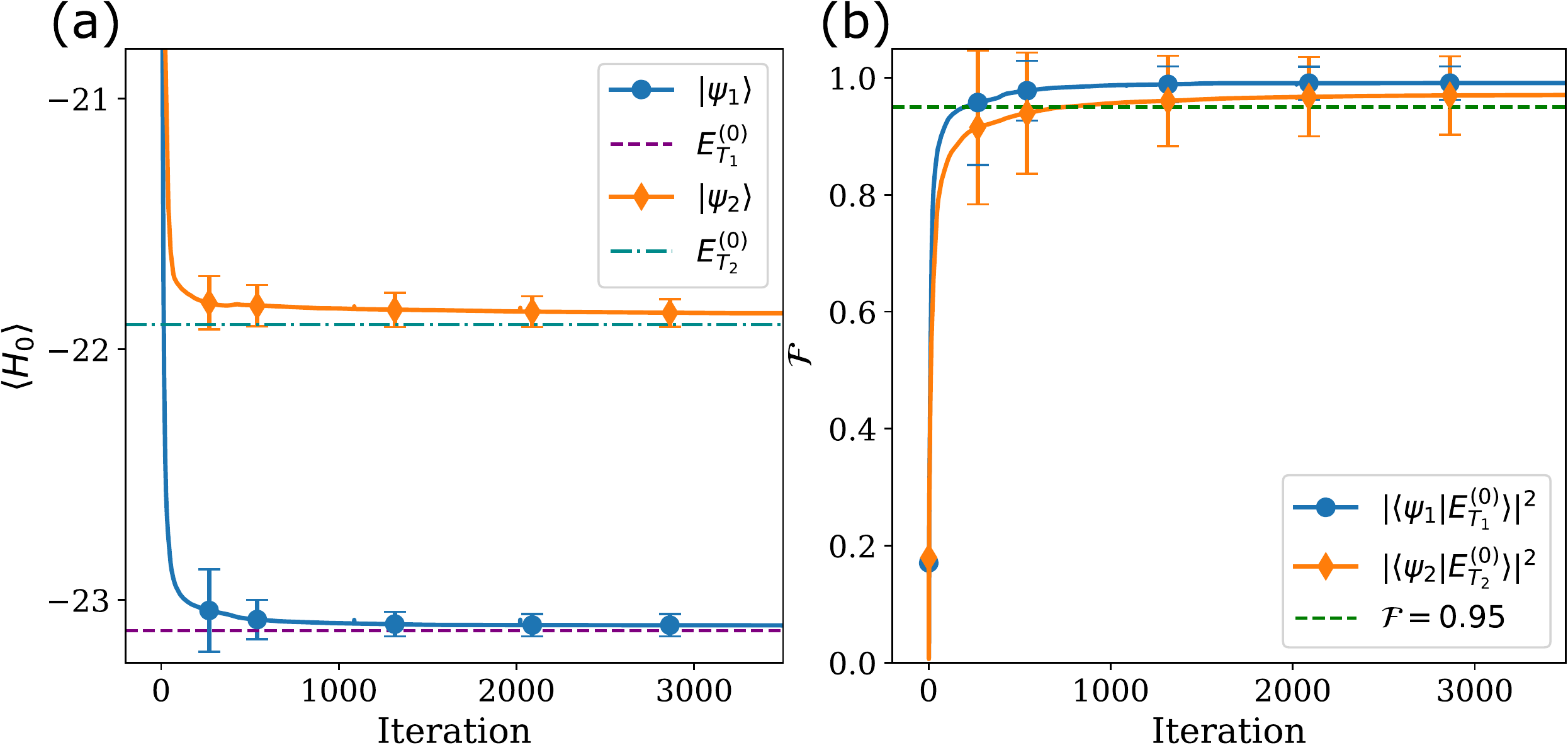}

\caption{The SSVQE simulation for simultaneously generating two eigenstates $\ket{E_{T_1}^{(0)}}$ and $\ket{E_{T_2}^{(0)}}$ using $20$ layers of $N=14$ $S_z$-conserving circuit. The average energy, shown in panel (a), and the corresponding obtainable fidelity, shown in panel (b), as a function of optimization iteration $n_I$. The error bars are computed through averaging over $100$ random initial samples.}
\label{fig:heis_e1-e2_result}
\end{figure}

For the preparation of $\ket{E_{T_1}^{(0)}}$ and $\ket{E_{T_2}^{(0)}}$ via SSVQE, the cost function is updated as 
\begin{equation} 
    \mathrm{cost} = \sum_i w_i ({\langle H_0 \rangle}_{i} + \beta {\langle (S_{tot}^{2} - 2)^2 \rangle}_{i})
    \label{cost_triplet}
\end{equation}
where, again, $\beta$ is a sufficiently large positive scalar. This type of cost functions are different from the one used for the singlets, see Eq.~\eqref{eq:e0+e3_cost}, as the operator in the penalty term requires four point correlation functions. In Ref.~\cite{kuroiwa2021penalty}, it is shown that $\beta$ for these type of cost function can be chosen to be smaller, lower bounded by the energy gap. In fact, for our triplet cost function in Eq.~\eqref{cost_triplet}, choosing $\beta=2$ is enough for convergence.
In Figs.~\ref{fig:heis_e1-e2_result}(a)-(b), we plot the average energy $\langle H_0 \rangle$ and the obtainable fidelity as a function of optimization iteration, respectively, for a circuit of system size $N=14$ with $20$ layers. Similar to the previous case, the results are repeated for $100$ random samples over which the error bars are estimated. In the numerical simulation, we set $\beta=2$. Indeed, the SSVQE successfully generates $\ket{E_{T_1}^{(0)}}$ and $\ket{E_{T_2}^{(0)}}$, simultaneously, achieving the fidelity $\mathcal{F}> 0.95$.
It is worth emphasizing that in comparison with the preparation of $\ket{E_{S_1}}$ and $\ket{E_{S_2}}$, targeting the triplet eigenstates is more difficult  in terms of required circuit layers. However, thanks to a stronger penalization term in the cost function, the error bars for triplets are smaller than their singlet counterparts, in particular at initial iterations, see Fig.~\ref{fig:heis_e0-e3_result}. 
Our procedure shows significant improvement over the results without symmetry, shown in Fig.~\ref{fig:hard-eff_heis_expt-vs-layer}(c), in which $\ket{E_{T_2}}$ fails to be generated for a fairly small system size of $N=10$ with even $25$ layers of hardware-efficient ansatz and $\sim 5000$ iterations. Similarly, one can target $\ket{E_{T_1}^{(+1)}}$ and $\ket{E_{T_2}^{(+1)}}$ (or equivalently $\ket{E_{T_1}^{(-1)}}$ and $\ket{E_{T_2}^{(-1)}}$) using the same circuit with proper choice of initial states with $s_z=+1$ ($s_z=-1$). For the sake of brevity, we do not present the results of these simulations.

\section{Hybrid versus Hardware symmetry preserving}

In the previous section, we showed how hybrid symmetry preserving method can enhance the performance of VQE through a combination of $S_z$-conserving circuit and updated cost function. In this section, we introduce hardware symmetry preserving approach in which both of the symmetries are included in the design of the circuit and thus the cost function only minimizes the average energy. In Fig.~\ref{fig:ansatz}(d), we present one layer of a quantum circuit which conserves the $S_{tot}$. This ansatz is very similar to $S_z$-conserving circuit except phase gates which are removed. As mentioned before, this circuit not only conserves the $z$ component of the spin, namely $s_z$, but also preserves the total spin $s$ as well. Thus, by choosing a proper initial state with a specific total spin $s$ and a given spin component $s_z$ one can guarantee the preservation of these symmetries in the output state $\ket{\psi(\vec{\theta})}$. For the case of global singlets, i.e. $s=0$, one simple initial state is $\ket{\psi_0}=\otimes^{N/2} \ket{\psi^-}$, where $\ket{\psi^-} = (\ket{01} - \ket{10})/\sqrt{2}$ is a two qubit singlet state. To generate this initial state, one has to use at least $N/2$ extra CNOTs at the beginning of the quantum circuit. In the SSVQE, if one wants to target two global singlets, i.e. $\ket{E_{S_1}}$ and $\ket{E_{S_2}}$ simultaneously, one has to generate another global singlet initial state which demands extra CNOTs. In fact, by increasing the number of target eigenstates, generating proper initial states which are all orthogonal and satisfy $s=0$ become more complex and demand extra CNOT gates. In the case of global triplet with $s=1$, the situation is simpler. For instance, by taking the quantum state $\otimes^{N/2} \ket{\psi^-}$ and locally converting one of the two-qubit singlets $\ket{\psi^-}$ into a two-qubit triplet (with a desire $s_z$), one can generate $N/2$ different orthogonal global triplet initial states. For higher total spins (i.e. $s>1$), the simple converting of two-qubit singlets into triplets does not create an initial state with a given $s$. Therefore, $S_{tot}$-conserving symmetry circuits become more complex, see Ref.~\cite{Gard2020} for more detailed discussion.  

\begin{figure}[h]
  \centering
  \includegraphics[width=\columnwidth]{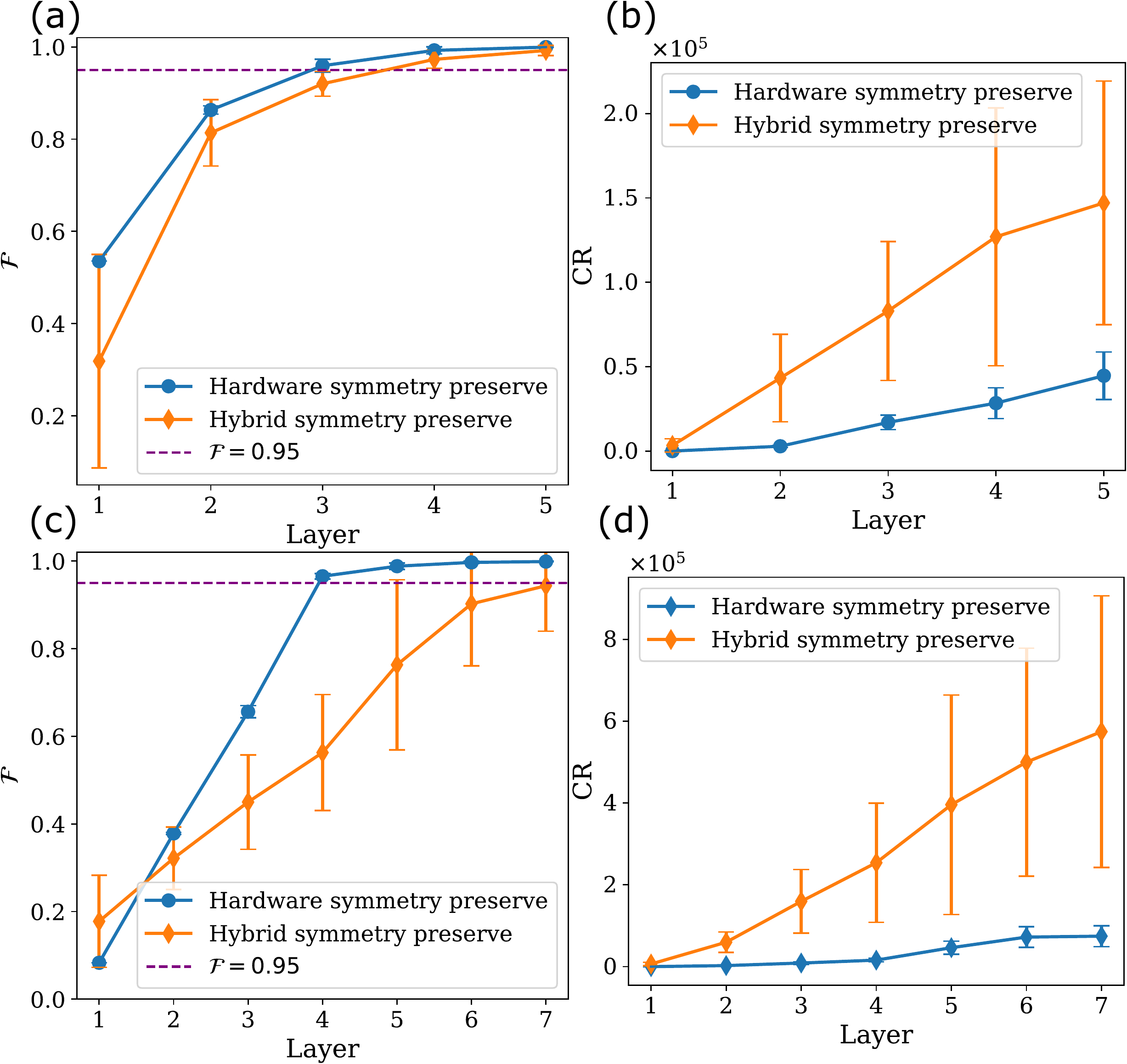}

\caption{Comparison between hardware and hybrid  symmetry preserving methods for simulating  the ground state  $\ket{E_{S_1}}$ and the first excited state $\ket{E_{T_1}^{(0)}}$ of a Heisenberg chain of system size $N=14$. For the simulation of $\ket{E_{S_1}}$, the fidelity, shown in panel (a), and the average classical resources $C_R$, shown in panel (b),  are plotted as a function of circuit layer using $S_z$- and $S_{tot}$-conserving circuits. Similarly, for the simulation of $\ket{E_{T_1}^{(0)}}$, the fidelity, shown in panel (c), and the average classical resources $C_R$, shown in panel (d),  are plotted as a function of circuit layer using $S_z$- and $S_{tot}$-conserving circuits. The error bars are computed through averaging over $50$ random initial samples.
}
\label{fig:symmetry_in_circ_vs_cost}
\end{figure}

To better understand the impact of symmetry on VQE simulation, we compare the performance of the hybrid symmetry preserving approach, which uses $S_{z}$-conserving circuit, and the hardware symmetry preserving approach, which utilizes $S_{tot}$-conserving circuits. As an example, we focus on generating the ground and the first excited state of the Heisenberg Hamiltonian, namely $\ket{E_{S_1}}$ (with $s=0$ and $s_z=0$) and $\ket{E_{T_1}^{(0)}}$ (with $s=1$ and $s_z=0$), with the size of $N=14$. In Fig.~\ref{fig:symmetry_in_circ_vs_cost}(a), we plot the obtainable fidelity for targeting the ground state $\ket{E_{S_1}}$ as a function of circuit layers for the two circuits. As the figure shows, the hardware symmetry preserving approach reaches the fidelity $\mathcal{F}>0.95$ with less layers. Despite requiring extra CNOTs for preparing its initial global singlet state $\ket{\psi_0}=\otimes^{N/2} \ket{\psi^-}$, the hardware symmetry preserving method is still more efficient in terms of two-qubit entangling gates, using $117$ CNOTs versus $156$ CNOTs in hybrid preserving symmetry method. In Fig.~\ref{fig:symmetry_in_circ_vs_cost}(b), we plot the classical resources, defined in Eq.~\eqref{CR_eq}, as a function of layers which shows significant advantage for the hardware symmetry preserving method. The superiority of the hardware symmetry preserving method becomes more evident when one targets the first excited state $\ket{E_{T_1}^{(0)}}$. In Fig.~\ref{fig:symmetry_in_circ_vs_cost}(c), we plot the fidelity as a function of layers for both quantum circuits. The hardware symmetry preserving method shows significant advantage as it reaches the fidelity $\mathcal{F}>0.95$ only after $4$ layers (with $156$ CNOTs) in comparison with $7$ layers of hybrid symmetry preserving method (with $273$ CNOTs). In Fig.~\ref{fig:symmetry_in_circ_vs_cost}(d), the corresponding classical resources are compared. Interestingly, the hardware symmetry preserving method not only demands much less classical resources, but also benefits from less fluctuations.  

For the sake of completeness, we also compare the performance of hybrid and hardware symmetry preserving methods using SSVQE for targeting $\ket{E_{S_1}}$ and $\ket{E_{S_2}}$ as well as $\ket{E_{T_1}^{(0)}}$ and $\ket{E_{T_2}^{(0)}}$ with the system size $N=14$. The results are given in Table.~\ref{tbl:hybrid_hardware_compare}. As the results show, hardware symmetry preserving method outperforms the hybrid symmetry preserving approach.

This analysis shows that the hardware symmetry preserving approach is the most efficient way for exploiting symmetries in SSVQE in terms of both quantum and classical resources. However, the price that one has to pay is the more complex circuit design. In particular, for total spin $s>1$ the initialization may indeed need complex quantum circuits~\cite{Gard2020}.

\setlength{\tabcolsep}{0.15em}
\begin{table*}[t]
    \centering
    \begin{tabular}{|c|c|c|c|c|c|c|c|c|c|c|}
        \hline
        & \multicolumn{5}{c|}{Hybrid Symmetry Preserving }& \multicolumn{5}{c|}{Hardware symmetry preserving} \\
        \hline
        Layer & 16 & 17 & 18 & 19 & 20 & 16 & 17 & 18 & 19 & 20 \\
        \hline 
        $\mathcal{F}_{\ket{E_{S_1}}}$ & 0.96$\pm$0.16 & 0.96$\pm$0.16 & 0.96$\pm$0.16 & 0.98$\pm$0.11 & 0.98$\pm$0.09 & 1.00$\pm$0.00 & 1.00$\pm$0.00 & 1.00$\pm$0.00 & 1.00$\pm$0.00 & 1.00$\pm$0.00 \\
        \hline
        $\mathcal{F}_{\ket{E_{S_2}}}$ & 0.91$\pm$0.21 & 0.95$\pm$0.19 & 0.95$\pm$0.15 & 0.97$\pm$0.14 & 0.97$\pm$0.13 & 0.95$\pm$0.03 & 0.96$\pm$0.02 & 0.98$\pm$0.02 & 0.99$\pm$0.01 & 0.99$\pm$0.01 \\
        \hline
        $\mathcal{F}_{\ket{E_{T_1}^{(0)}}}$ & 0.97$\pm$0.06 & 0.97$\pm$0.06 & 0.98$\pm$0.03 & 0.99$\pm$0.03 & 0.99$\pm$0.01 & 1.00$\pm$0.00 & 1.00$\pm$0.00 & 1.00$\pm$0.00 & 1.00$\pm$0.00 & 1.00$\pm$0.00\\
        \hline
        $\mathcal{F}_{\ket{E_{T_2}^{(0)}}}$ & 0.91$\pm$0.12 & 0.95$\pm$0.07 & 0.97$\pm$0.07 & 0.97$\pm$0.05 & 0.99$\pm$0.02 & 1.00$\pm$0.00 & 1.00$\pm$0.00 & 1.00$\pm$0.00 & 1.00$\pm$0.00 & 1.00$\pm$0.00\\
        \hline
    \end{tabular}
    \caption{The achievable fidelity as a function of circuit layer for both hybrid and hardware symmetry preserving methods using SSVQE for simultaneously targeting $\ket{E_{S_1}}$ and $\ket{E_{S_2}}$, as well as $\ket{E_{T_1}^{(0)}}$ and $\ket{E_{T_2}^{(0)}}$. The table shows that hardware symmetry preserving method outperforms the hybrid symmetry preserving approach as higher fidelities can be achieved for the same number of layers. }
    \label{tbl:hybrid_hardware_compare}
\end{table*}

\section{Generality of symmetry method}

So far, we have focused on Heisenberg Hamiltonian for which we exploit $S_{z}$ and $S_{tot}$ symmetry for preparing several low-energy eigenstates. Here, we show that symmetry method can be generalized to other Hamiltonians with different symmetries. In particular, we will show that exploiting symmetries can also enhance the VQE performance for simulating free fermionic systems. For example, we consider the Ising Hamiltonian with transverse field 
\begin{equation}
    H_{I} = J_z \sum_{i = 1}^{N - 1} \sigma_z^i \sigma_z^{i+1} + h_x \sum_{i=1}^{N} \sigma_x^i 
    \label{Ising_ham}
\end{equation}
where $J_z$ is the exchange coupling and $h_x$ is the strength of the transverse magnetic field. It is well known that this Hamiltonian has a quantum phase transition at $J_z / h_x = 1$. At the critical point where the quantum phase transition takes place, the ground state is highly entangled and has a complex form. Indeed, it has been shown that at the critical point the convergence of VQE requires more quantum and classical resources~\cite{Chufan2022longrange}. In other words, one needs a deeper quantum circuit as well as more iterations to reach the target state. 
Therefore, we focus at the critical point, as the most complex point in the phase diagram, for generating the first two eigenstates of the Hamiltonian. The Ising Hamiltonian commutes with the total spin flip operator $\widetilde{S_x}=\prod_{i=1}^N \sigma_x^i$, namely $[H_{I}, \widetilde{S_x}]=0$. This implies that the $H_{Ising}$ and $\widetilde{S_x}$ have common eigenvectors $\ket{E_k}$. Since the eigenvalues of $\widetilde{S_x}$ are $\pm 1$, then $\langle E_k | \widetilde{S_x} | E_k  \rangle = \pm 1$ for every eigenstate $\ket{E_k}$. In particular, for even $N$, one can verify that the ground state and the first excited state satisfy $\langle E_1 | \widetilde{S_x} | E_1  \rangle=1$ and $\langle E_2 | \widetilde{S_x} | E_2  \rangle=-1$, respectively. 

\begin{figure}[t]
  \centering
  \includegraphics[width=\columnwidth]{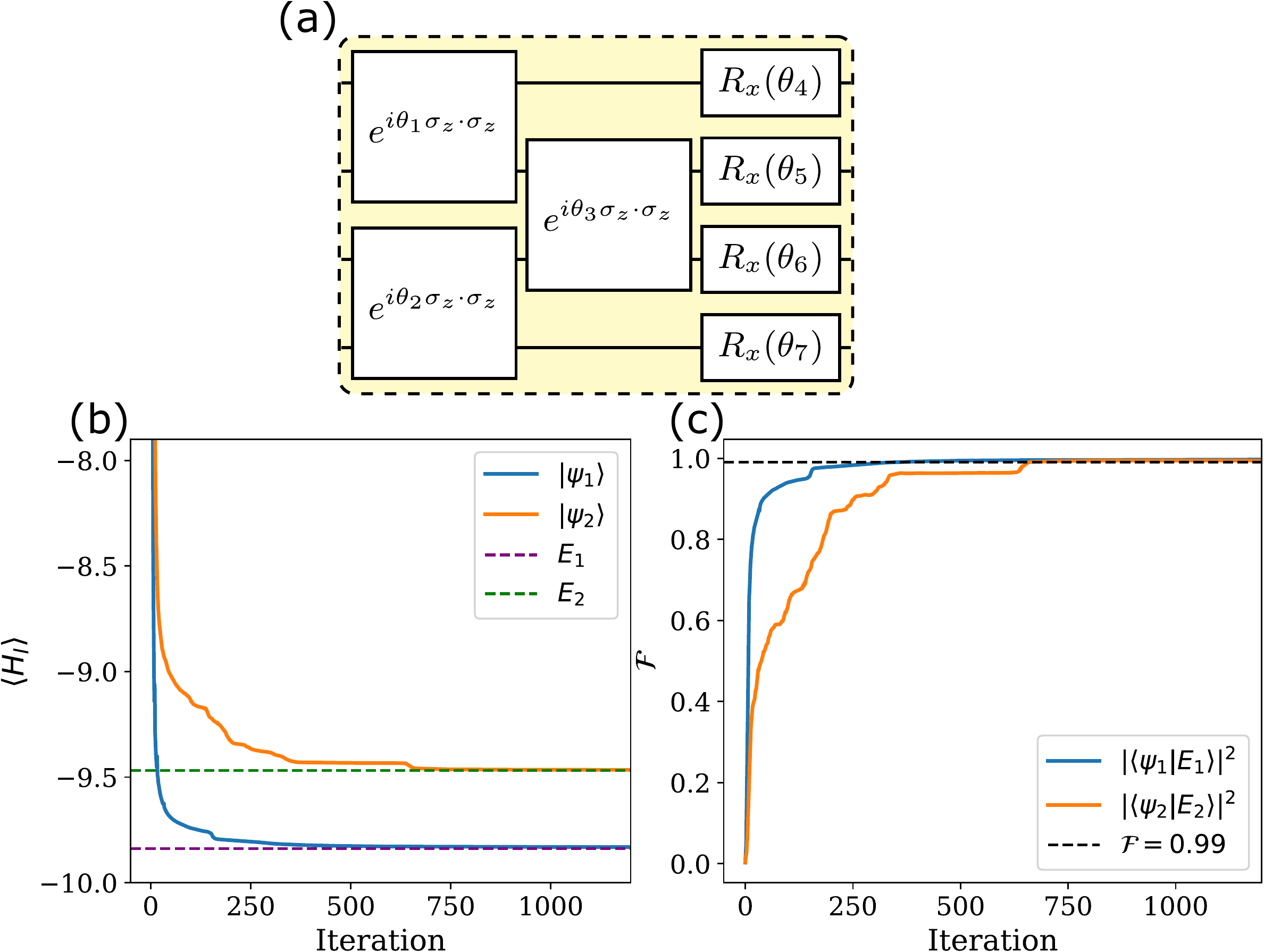}

\caption{ VQE simulation of the Ising model in transverse field. (a) A single layer circuit of the ansatz for Ising Hamiltonian of system size $N=4$. (b) The average energy as a function of optimization iteration $n_I$ for the ground state $\ket{E_1}$ and the first excited state $\ket{E_2}$ for a system of size $N=8$ on a circuit with  $4$ layers. (c) The corresponding average fidelity for $\ket{E_1}$ and $\ket{E_2}$ as a function of optimization iteration $n_I$.}
\label{fig:ising_result}
\end{figure}

Since the symmetry operator $\widetilde{S_x}$ is a global action, designing a quantum circuit that implements this symmetry is very difficult. Therefore, hardware symmetry preserving approach becomes infeasible. Consequently, we use updated cost function for generating the first two eigenstates of the Hamiltonian. We use the circuit shown in Fig.~\ref{fig:ising_result}(a).
For starting the state preparation, we initialize the circuit in the quantum state $\otimes^{N} \ket{+}$, where $\ket{+}=(\ket{0} + \ket{1})/\sqrt{2}$. While for generating $\ket{E_1}$ a simple cost function $\langle H_{I} \rangle$ is enough, for preparation of $\ket{E_2}$ one can simply update the cost function as 
\begin{equation}
    \mathrm{cost} = \langle H_I \rangle + \beta (\langle \widetilde{S_x} \rangle + 1)^2
\end{equation}
where $\beta$ is a positive scalar to make the two terms of the same order (here, we put $\beta=1$). In Figs.~\ref{fig:ising_result}(b)-(c), we plot the average energy $\langle H_{I} \rangle$ and the average fidelity over $50$ random samples as a function of optimization iterations, respectively, for a system of size $N=8$ with $4$ layers of ansatz. As shown in the plot, as optimization iteration increases, the VQE successfully generate $\ket{E_2}$ with fidelity $\mathcal{F} > 0.99$. For generating more eigenstates one has to combine the symmetries with SSVQE algorithm as we did above for the Heisenberg Hamiltonian.

\section{Conclusion}

The VQE has emerged as one of the leading NISQ algorithms with the potential of achieving quantum advantage. So far, it has been successfully applied for simulating the ground state of condensed matter systems and several chemical materials. Nonetheless, generalization of the VQE to higher energy eigenstates is very resource consuming, demanding deep circuits and lengthy classical optimizations. In this paper, we address this problem by exploiting symmetries to enhance both quantum and classical resource efficiencies of the VQE algorithm. The symmetry enhanced resource efficiency becomes even more effective when higher energy eigenstates are targeted. Indeed, some of the excited states cannot be reached without including symmetries in the VQE algorithm. We have considered two ways for incorporating symmetries. In the first approach, which we call it hardware symmetry preserving method, all the symmetries are included in the circuit. In the second method, the symmetries are integrated in the cost function. Our results show that the hardware symmetry preserving method significantly outperforms the penalization of the cost function. However, implementing all symmetries in the design of the circuit may not be practical. Therefore, we have introduced the hybrid symmetry preserving method in which some of the symmetries are included in the circuit and the rest are incorporated in the cost function. This allows to simultaneously improve the resource efficiency and keep the circuit design simple. Our proposal achieves significant resource efficiency and thus paves the way for achieving quantum advantage on NISQ simulators. In addition, it is very timely and can be adopted in existing quantum simulators.

Note that our method is applicable only we have prior knowledge about the symmetries of the system. For any given Hamiltonian, in order to investigate the symmetries of the system, one can use a small system as symmetries are not length dependent. In addition, usually the symmetry values of the low energy eigenstates can also be determined for small sizes and directly be used for large systems. In other words, the fact that symmetries are not length dependent can be very beneficial in specifying the symmetries and incorporating them in VQE algorithms.

\section*{Data availability}
All the source codes for generating the data is available~\cite{Chufan2022symcode}. In addition, the data which have been for generating plots can be provided upon reasonable request from the authors. 

\section*{Acknowledgments}
A.B. acknowledges support from the National Key
R\&D Program of China (Grant No.2018YFA0306703),
the National Science Foundation of China (Grants
No. 12050410253, No. 92065115 and No. 12274059), and the Ministry of Science and Technology of China (Grant No.
QNJ2021167001L).

\bibliographystyle{quantum}
\bibliography{vqe_symmetry_enhanced_bib}

\begin{thebibliography}{100}

\bibitem{kokail2019self}
Christian Kokail, Christine Maier, Rick van Bijnen, Tiff Brydges, Manoj~K
  Joshi, Petar Jurcevic, Christine~A Muschik, Pietro Silvi, Rainer Blatt,
  Christian~F Roos, et~al.
\newblock ``Self-verifying variational quantum simulation of lattice models''.
\newblock \href{https://dx.doi.org/10.1038/s41586-019-1177-4}{Nature {\bf 569},
  355--360}~(2019).

\bibitem{aspuru2005simulated}
Al{\'a}n Aspuru-Guzik, Anthony~D Dutoi, Peter~J Love, and Martin Head-Gordon.
\newblock ``Simulated quantum computation of molecular energies''.
\newblock \href{https://dx.doi.org/10.1126/science.1113479}{Science {\bf 309},
  1704--1707}~(2005).

\bibitem{helgaker2014molecular}
Trygve Helgaker, Poul Jorgensen, and Jeppe Olsen.
\newblock ``Molecular electronic-structure theory''.
\newblock \href{https://dx.doi.org/10.1002/9781119019572}{John Wiley \& Sons,
  Ltd}. ~(2013).

\bibitem{orus2019quantum}
Roman Orus, Samuel Mugel, and Enrique Lizaso.
\newblock ``Quantum computing for finance: Overview and prospects''.
\newblock \href{https://dx.doi.org/10.1016/j.revip.2019.100028}{Reviews in
  Physics {\bf 4}, 100028}~(2019).

\bibitem{rebentrost2018quantum}
Patrick Rebentrost, Brajesh Gupt, and Thomas~R Bromley.
\newblock ``Quantum computational finance: Monte carlo pricing of financial
  derivatives''.
\newblock \href{https://dx.doi.org/10.1103/physreva.98.022321}{Phys. Rev. A
  {\bf 98}, 022321}~(2018).

\bibitem{egger2020quantum}
Daniel~J Egger, Claudio Gambella, Jakub Marecek, Scott McFaddin, Martin
  Mevissen, Rudy Raymond, Andrea Simonetto, Stefan Woerner, and Elena Yndurain.
\newblock ``Quantum computing for finance: state of the art and future
  prospects''.
\newblock \href{https://dx.doi.org/10.1109/tqe.2020.3030314}{IEEE Transactions
  on Quantum Engineering}~(2020).

\bibitem{bordia2017probing}
Pranjal Bordia, Henrik L{\"u}schen, Sebastian Scherg, Sarang Gopalakrishnan,
  Michael Knap, Ulrich Schneider, and Immanuel Bloch.
\newblock ``Probing slow relaxation and many-body localization in
  two-dimensional quasiperiodic systems''.
\newblock \href{https://dx.doi.org/10.1103/physrevx.7.041047}{Phys. Rev. X {\bf
  7}, 041047}~(2017).

\bibitem{schreiber2015observation}
Michael Schreiber, Sean~S Hodgman, Pranjal Bordia, Henrik~P L{\"u}schen, Mark~H
  Fischer, Ronen Vosk, Ehud Altman, Ulrich Schneider, and Immanuel Bloch.
\newblock ``Observation of many-body localization of interacting fermions in a
  quasirandom optical lattice''.
\newblock \href{https://dx.doi.org/10.1126/science.aaa7432}{Science {\bf 349},
  842--845}~(2015).

\bibitem{gross2017quantum}
Christian Gross and Immanuel Bloch.
\newblock ``Quantum simulations with ultracold atoms in optical lattices''.
\newblock \href{https://dx.doi.org/10.1126/science.aal3837}{Science {\bf 357},
  995--1001}~(2017).

\bibitem{hempel2018quantum}
Cornelius Hempel, Christine Maier, Jonathan Romero, Jarrod McClean, Thomas
  Monz, Heng Shen, Petar Jurcevic, Ben~P Lanyon, Peter Love, Ryan Babbush,
  et~al.
\newblock ``Quantum chemistry calculations on a trapped-ion quantum
  simulator''.
\newblock \href{https://dx.doi.org/10.1103/PhysRevX.8.031022}{Phys. Rev. X {\bf
  8}, 031022}~(2018).

\bibitem{lanyon2011universal}
Ben~P Lanyon, Cornelius Hempel, Daniel Nigg, Markus M{\"u}ller, Rene Gerritsma,
  F~Z{\"a}hringer, Philipp Schindler, Julio~T Barreiro, Markus Rambach, Gerhard
  Kirchmair, et~al.
\newblock ``Universal digital quantum simulation with trapped ions''.
\newblock \href{https://dx.doi.org/10.1126/science.1208001}{Science {\bf 334},
  57--61}~(2011).

\bibitem{aspuru2012photonic}
Al{\'a}n Aspuru-Guzik and Philip Walther.
\newblock ``Photonic quantum simulators''.
\newblock \href{https://dx.doi.org/10.1038/nphys2253}{Nat. Phys. {\bf 8},
  285--291}~(2012).

\bibitem{wang2020integrated}
Jianwei Wang, Fabio Sciarrino, Anthony Laing, and Mark~G Thompson.
\newblock ``Integrated photonic quantum technologies''.
\newblock \href{https://dx.doi.org/10.1038/s41566-019-0532-1}{Nat. Photonics
  {\bf 14}, 273--284}~(2020).

\bibitem{hensgens2017quantum}
Toivo Hensgens, Takafumi Fujita, Laurens Janssen, Xiao Li, CJ~Van~Diepen,
  Christian Reichl, Werner Wegscheider, S~Das Sarma, and Lieven~MK Vandersypen.
\newblock ``Quantum simulation of a fermi--hubbard model using a semiconductor
  quantum dot array''.
\newblock \href{https://dx.doi.org/10.1038/nature23022}{Nature {\bf 548},
  70--73}~(2017).

\bibitem{salfi2016quantum}
J~Salfi, JA~Mol, R~Rahman, G~Klimeck, MY~Simmons, LCL Hollenberg, and S~Rogge.
\newblock ``Quantum simulation of the hubbard model with dopant atoms in
  silicon''.
\newblock \href{https://dx.doi.org/10.1038/ncomms11342}{Nat. Commun. {\bf 7},
  1--6}~(2016).

\bibitem{arute2020hartree}
Frank Arute, Kunal Arya, Ryan Babbush, Dave Bacon, Joseph~C Bardin, Rami
  Barends, Sergio Boixo, Michael Broughton, Bob~B Buckley, David~A Buell,
  et~al.
\newblock ``Hartree-fock on a superconducting qubit quantum computer''.
\newblock \href{https://dx.doi.org/10.1126/science.abb9811}{Science {\bf 369},
  1084--1089}~(2020).

\bibitem{barends2016digitized}
Rami Barends, Alireza Shabani, Lucas Lamata, Julian Kelly, Antonio Mezzacapo,
  Urtzi Las~Heras, Ryan Babbush, Austin~G Fowler, Brooks Campbell, Yu~Chen,
  et~al.
\newblock ``Digitized adiabatic quantum computing with a superconducting
  circuit''.
\newblock \href{https://dx.doi.org/10.1038/nature17658}{Nature {\bf 534},
  222--226}~(2016).

\bibitem{preskill2018quantum}
John Preskill.
\newblock ``Quantum computing in the nisq era and beyond''.
\newblock \href{https://dx.doi.org/10.22331/q-2018-08-06-79}{Quantum {\bf 2},
  79}~(2018).

\bibitem{bharti2021noisy}
Kishor Bharti, Alba Cervera-Lierta, Thi~Ha Kyaw, Tobias Haug, Sumner
  Alperin-Lea, Abhinav Anand, Matthias Degroote, Hermanni Heimonen, Jakob~S.
  Kottmann, Tim Menke, Wai-Keong Mok, Sukin Sim, Leong-Chuan Kwek, and Alán
  Aspuru-Guzik.
\newblock ``Noisy intermediate-scale quantum algorithms''.
\newblock \href{https://dx.doi.org/10.1103/revmodphys.94.015004}{Rev. Mod.
  Phys.{\bf 94}}~(2022).

\bibitem{peruzzo2014variational}
Alberto Peruzzo, Jarrod McClean, Peter Shadbolt, Man-Hong Yung, Xiao-Qi Zhou,
  Peter~J Love, Al{\'a}n Aspuru-Guzik, and Jeremy~L O’brien.
\newblock ``A variational eigenvalue solver on a photonic quantum processor''.
\newblock \href{https://dx.doi.org/10.1038/ncomms5213}{Nat. Commun. {\bf 5},
  1--7}~(2014).

\bibitem{cerezo2021variational}
Marco Cerezo, Andrew Arrasmith, Ryan Babbush, Simon~C Benjamin, Suguru Endo,
  Keisuke Fujii, Jarrod~R McClean, Kosuke Mitarai, Xiao Yuan, Lukasz Cincio,
  et~al.
\newblock ``Variational quantum algorithms''.
\newblock \href{https://dx.doi.org/10.1038/s42254-021-00348-9}{Nat. Rev.
  Phys.Pages 1--20}~(2021).

\bibitem{mcclean2016theory}
Jarrod~R McClean, Jonathan Romero, Ryan Babbush, and Al{\'a}n Aspuru-Guzik.
\newblock ``The theory of variational hybrid quantum-classical algorithms''.
\newblock \href{https://dx.doi.org/10.1088/1367-2630/18/2/023023}{New J. Phys.
  {\bf 18}, 023023}~(2016).

\bibitem{yuan2019theory}
Xiao Yuan, Suguru Endo, Qi~Zhao, Ying Li, and Simon~C Benjamin.
\newblock ``Theory of variational quantum simulation''.
\newblock \href{https://dx.doi.org/10.22331/q-2019-10-07-191}{Quantum {\bf 3},
  191}~(2019).

\bibitem{xin2020quantum}
Tao Xin, Xinfang Nie, Xiangyu Kong, Jingwei Wen, Dawei Lu, and Jun Li.
\newblock ``Quantum pure state tomography via variational hybrid
  quantum-classical method''.
\newblock \href{https://dx.doi.org/10.1103/PhysRevApplied.13.024013}{Phys. Rev.
  Applied {\bf 13}, 024013}~(2020).

\bibitem{biamonte2017quantum}
Jacob Biamonte, Peter Wittek, Nicola Pancotti, Patrick Rebentrost, Nathan
  Wiebe, and Seth Lloyd.
\newblock ``Quantum machine learning''.
\newblock \href{https://dx.doi.org/10.1038/nature23474}{Nature {\bf 549},
  195--202}~(2017).

\bibitem{arunachalam2017survey}
Srinivasan Arunachalam and Ronald de~Wolf.
\newblock ``A survey of quantum learning theory''~(2017).
\newblock  \href{http://arxiv.org/abs/1701.06806}{arXiv:1701.06806}.

\bibitem{ciliberto2018quantum}
Carlo Ciliberto, Mark Herbster, Alessandro~Davide Ialongo, Massimiliano Pontil,
  Andrea Rocchetto, Simone Severini, and Leonard Wossnig.
\newblock ``Quantum machine learning: a classical perspective''.
\newblock \href{https://dx.doi.org/10.1098/rspa.2017.0551}{Proceedings of the
  Royal Society A: Mathematical, Physical and Engineering Sciences {\bf 474},
  20170551}~(2018).

\bibitem{dunjko2018machine}
Vedran Dunjko and Hans~J Briegel.
\newblock ``Machine learning \& artificial intelligence in the quantum domain:
  a review of recent progress''.
\newblock \href{https://dx.doi.org/10.1088/1361-6633/aab406}{Reports on
  Progress in Physics {\bf 81}, 074001}~(2018).

\bibitem{farhi2018classification}
Edward Farhi and Hartmut Neven.
\newblock ``Classification with quantum neural networks on near term
  processors''~(2018).
\newblock  \href{http://arxiv.org/abs/1802.06002}{arXiv:1802.06002}.

\bibitem{schuld2019quantum}
Maria Schuld and Nathan Killoran.
\newblock ``Quantum machine learning in feature hilbert spaces''.
\newblock \href{https://dx.doi.org/10.1103/physrevlett.122.040504}{Phys. Rev.
  Lett. {\bf 122}, 040504}~(2019).

\bibitem{farhi2014quantum}
Edward Farhi, Jeffrey Goldstone, and Sam Gutmann.
\newblock ``A quantum approximate optimization algorithm''~(2014).
\newblock  \href{http://arxiv.org/abs/1411.4028}{arXiv:1411.4028}.

\bibitem{bravyi2020obstacles}
Sergey Bravyi, Alexander Kliesch, Robert Koenig, and Eugene Tang.
\newblock ``Obstacles to variational quantum optimization from symmetry
  protection''.
\newblock \href{https://dx.doi.org/10.1103/physrevlett.125.260505}{Phys. Rev.
  Lett. {\bf 125}, 260505}~(2020).

\bibitem{cirstoiu2020variational}
Cristina Cirstoiu, Zoe Holmes, Joseph Iosue, Lukasz Cincio, Patrick~J Coles,
  and Andrew Sornborger.
\newblock ``Variational fast forwarding for quantum simulation beyond the
  coherence time''.
\newblock \href{https://dx.doi.org/10.1038/s41534-020-00302-0}{Npj Quantum Inf.
  {\bf 6}, 1--10}~(2020).

\bibitem{gibbs2021longtime}
Joe Gibbs, Kaitlin Gili, Zoë Holmes, Benjamin Commeau, Andrew Arrasmith,
  Lukasz Cincio, Patrick~J. Coles, and Andrew Sornborger.
\newblock ``Long-time simulations with high fidelity on quantum
  hardware''~(2021).
\newblock  \href{http://arxiv.org/abs/2102.04313}{arXiv:2102.04313}.

\bibitem{mcardle2019variational}
Sam McArdle, Tyson Jones, Suguru Endo, Ying Li, Simon~C Benjamin, and Xiao
  Yuan.
\newblock ``Variational ansatz-based quantum simulation of imaginary time
  evolution''.
\newblock \href{https://dx.doi.org/10.1038/s41534-019-0187-2}{Npj Quantum Inf.
  {\bf 5}, 1--6}~(2019).

\bibitem{heya2019subspace}
Kentaro Heya, Ken~M Nakanishi, Kosuke Mitarai, and Keisuke Fujii.
\newblock ``Subspace variational quantum simulator''~(2019).
\newblock  \href{http://arxiv.org/abs/1904.08566}{arXiv:1904.08566}.

\bibitem{huh2014linear}
Joonsuk Huh, Sarah Mostame, Takatoshi Fujita, Man-Hong Yung, and Al{\'a}n
  Aspuru-Guzik.
\newblock ``Linear-algebraic bath transformation for simulating complex open
  quantum systems''.
\newblock \href{https://dx.doi.org/10.1088/1367-2630/16/12/123008}{New J. Phys.
  {\bf 16}, 123008}~(2014).

\bibitem{hu2020quantum}
Zixuan Hu, Rongxin Xia, and Sabre Kais.
\newblock ``A quantum algorithm for evolving open quantum dynamics on quantum
  computing devices''.
\newblock \href{https://dx.doi.org/10.1038/s41598-020-60321-x}{Sci. Rep. {\bf
  10}, 1--9}~(2020).

\bibitem{endo2020variational}
Suguru Endo, Jinzhao Sun, Ying Li, Simon~C Benjamin, and Xiao Yuan.
\newblock ``Variational quantum simulation of general processes''.
\newblock \href{https://dx.doi.org/10.1103/physrevlett.125.010501}{Phys. Rev.
  Lett. {\bf 125}, 010501}~(2020).

\bibitem{haug2020generalized}
Tobias Haug and Kishor Bharti.
\newblock ``Generalized quantum assisted simulator''~(2020).
\newblock  \href{http://arxiv.org/abs/2011.14737}{arXiv:2011.14737}.

\bibitem{meyer2021variational}
Johannes~Jakob Meyer, Johannes Borregaard, and Jens Eisert.
\newblock ``A variational toolbox for quantum multi-parameter estimation''.
\newblock \href{https://dx.doi.org/10.1038/s41534-021-00425-y}{Npj Quantum Inf.
  {\bf 7}, 1--5}~(2021).

\bibitem{meyer2021fisher}
Johannes~Jakob Meyer.
\newblock ``Fisher information in noisy intermediate-scale quantum
  applications''.
\newblock \href{https://dx.doi.org/10.22331/q-2021-09-09-539}{Quantum {\bf 5},
  539}~(2021).

\bibitem{beckey2020variational}
Jacob~L. Beckey, M.~Cerezo, Akira Sone, and Patrick~J. Coles.
\newblock ``Variational quantum algorithm for estimating the quantum fisher
  information''.
\newblock \href{https://dx.doi.org/10.1103/physrevresearch.4.013083}{Phys. Rev.
  Res.{\bf 4}}~(2022).

\bibitem{kaubruegger2019variational}
Raphael Kaubruegger, Pietro Silvi, Christian Kokail, Rick van Bijnen, Ana~Maria
  Rey, Jun Ye, Adam~M Kaufman, and Peter Zoller.
\newblock ``Variational spin-squeezing algorithms on programmable quantum
  sensors''.
\newblock \href{https://dx.doi.org/10.1103/physrevlett.123.260505}{Phys. Rev.
  Lett. {\bf 123}, 260505}~(2019).

\bibitem{koczor2020variational}
B{\'a}lint Koczor, Suguru Endo, Tyson Jones, Yuichiro Matsuzaki, and Simon~C
  Benjamin.
\newblock ``Variational-state quantum metrology''.
\newblock \href{https://dx.doi.org/10.1088/1367-2630/ab965e}{New J. Phys. {\bf
  22}, 083038}~(2020).

\bibitem{ma2021adaptive}
Ziqi Ma, Pranav Gokhale, Tian-Xing Zheng, Sisi Zhou, Xiaofei Yu, Liang Jiang,
  Peter Maurer, and Frederic~T. Chong.
\newblock ``Adaptive circuit learning for quantum metrology''.
\newblock In 2021 {IEEE} International Conference on Quantum Computing and
  Engineering ({QCE}).
\newblock {IEEE}~(2021).

\bibitem{haug2021natural}
Tobias Haug and M.~S. Kim.
\newblock ``Natural parametrized quantum circuit''.
\newblock \href{https://dx.doi.org/10.1103/PhysRevA.106.052611}{Phys. Rev. A
  {\bf 106}, 052611}~(2022).

\bibitem{cao2021larger}
Changsu Cao, Jiaqi Hu, Wengang Zhang, Xusheng Xu, Dechin Chen, Fan Yu, Jun Li,
  Hanshi Hu, Dingshun Lv, and Man-Hong Yung.
\newblock ``Towards a larger molecular simulation on the quantum computer: Up
  to 28 qubits systems accelerated by point group symmetry''~(2021).
\newblock  \href{http://arxiv.org/abs/2109.02110}{arXiv:2109.02110}.

\bibitem{kandala2017hardware}
Abhinav Kandala, Antonio Mezzacapo, Kristan Temme, Maika Takita, Markus Brink,
  Jerry~M Chow, and Jay~M Gambetta.
\newblock ``Hardware-efficient variational quantum eigensolver for small
  molecules and quantum magnets''.
\newblock \href{https://dx.doi.org/10.1038/nature23879}{Nature {\bf 549},
  242--246}~(2017).

\bibitem{nam2020ground}
Yunseong Nam, Jwo-Sy Chen, Neal~C Pisenti, Kenneth Wright, Conor Delaney,
  Dmitri Maslov, Kenneth~R Brown, Stewart Allen, Jason~M Amini, Joel Apisdorf,
  et~al.
\newblock ``Ground-state energy estimation of the water molecule on a
  trapped-ion quantum computer''.
\newblock \href{https://dx.doi.org/10.1038/s41534-020-0259-3}{Npj Quantum Inf.
  {\bf 6}, 1--6}~(2020).

\bibitem{BravoPrieto2020scalingof}
Carlos Bravo-Prieto, Josep Lumbreras-Zarapico, Luca Tagliacozzo, and
  Jos{\'{e}}~I. Latorre.
\newblock ``Scaling of variational quantum circuit depth for condensed matter
  systems''.
\newblock \href{https://dx.doi.org/10.22331/q-2020-05-28-272}{{Quantum} {\bf
  4}, 272}~(2020).

\bibitem{Lyu2020accelerated}
Chufan Lyu, Victor Montenegro, and Abolfazl Bayat.
\newblock ``Accelerated variational algorithms for digital quantum simulation
  of many-body ground states''.
\newblock \href{https://dx.doi.org/10.22331/q-2020-09-16-324}{{Quantum} {\bf
  4}, 324}~(2020).

\bibitem{uvarov2020variational}
Alexey Uvarov, Jacob~D Biamonte, and Dmitry Yudin.
\newblock ``Variational quantum eigensolver for frustrated quantum systems''.
\newblock \href{https://dx.doi.org/10.1103/physrevb.102.075104}{Phys. Rev. B
  {\bf 102}, 075104}~(2020).

\bibitem{okada2022identification}
Ken~N. Okada, Keita Osaki, Kosuke Mitarai, and Keisuke Fujii.
\newblock ``Identification of topological phases using classically-optimized
  variational quantum eigensolver''~(2022).
\newblock  \href{http://arxiv.org/abs/2202.02909}{arXiv:2202.02909}.

\bibitem{chen2020demonstration}
Ming-Cheng Chen, Ming Gong, Xiaosi Xu, Xiao Yuan, Jian-Wen Wang, Can Wang,
  Chong Ying, Jin Lin, Yu~Xu, Yulin Wu, et~al.
\newblock ``Demonstration of adiabatic variational quantum computing with a
  superconducting quantum coprocessor''.
\newblock \href{https://dx.doi.org/10.1103/physrevlett.125.180501}{Phys. Rev.
  Lett. {\bf 125}, 180501}~(2020).

\bibitem{harrigan2021quantum}
Matthew~P Harrigan, Kevin~J Sung, Matthew Neeley, Kevin~J Satzinger, Frank
  Arute, Kunal Arya, Juan Atalaya, Joseph~C Bardin, Rami Barends, Sergio Boixo,
  et~al.
\newblock ``Quantum approximate optimization of non-planar graph problems on a
  planar superconducting processor''.
\newblock \href{https://dx.doi.org/10.1038/s41567-020-01105-y}{Nat. Phys. {\bf
  17}, 332--336}~(2021).

\bibitem{pagano2020quantum}
Guido Pagano, Aniruddha Bapat, Patrick Becker, Katherine~S Collins, Arinjoy De,
  Paul~W Hess, Harvey~B Kaplan, Antonis Kyprianidis, Wen~Lin Tan, Christopher
  Baldwin, et~al.
\newblock ``Quantum approximate optimization of the long-range ising model with
  a trapped-ion quantum simulator''.
\newblock \href{https://dx.doi.org/10.1073/pnas.2006373117}{Proceedings of the
  National Academy of Sciences {\bf 117}, 25396--25401}~(2020).

\bibitem{zhao2020measurement}
Andrew Zhao, Andrew Tranter, William~M Kirby, Shu~Fay Ung, Akimasa Miyake, and
  Peter~J Love.
\newblock ``Measurement reduction in variational quantum algorithms''.
\newblock \href{https://dx.doi.org/10.1103/physreva.101.062322}{Phys. Rev. A
  {\bf 101}, 062322}~(2020).

\bibitem{izmaylov2019unitary}
Artur~F Izmaylov, Tzu-Ching Yen, Robert~A Lang, and Vladyslav Verteletskyi.
\newblock ``Unitary partitioning approach to the measurement problem in the
  variational quantum eigensolver method''.
\newblock \href{https://dx.doi.org/10.1021/acs.jctc.9b00791}{J. Chem. Theory
  Comput. {\bf 16}, 190--195}~(2019).

\bibitem{verteletskyi2020measurement}
Vladyslav Verteletskyi, Tzu-Ching Yen, and Artur~F Izmaylov.
\newblock ``Measurement optimization in the variational quantum eigensolver
  using a minimum clique cover''.
\newblock \href{https://dx.doi.org/10.1063/1.5141458}{J. Chem. Phys. {\bf 152},
  124114}~(2020).

\bibitem{gokhale2020n}
Pranav Gokhale, Olivia Angiuli, Yongshan Ding, Kaiwen Gui, Teague Tomesh,
  Martin Suchara, Margaret Martonosi, and Frederic~T. Chong.
\newblock ``$o(n^3)$ measurement cost for variational quantum eigensolver on
  molecular hamiltonians''.
\newblock \href{https://dx.doi.org/10.1109/TQE.2020.3035814}{IEEE Transactions
  on Quantum Engineering {\bf 1}, 1--24}~(2020).

\bibitem{ralli2021implementation}
Alexis Ralli, Peter~J Love, Andrew Tranter, and Peter~V Coveney.
\newblock ``Implementation of measurement reduction for the variational quantum
  eigensolver''.
\newblock \href{https://dx.doi.org/10.1103/physrevresearch.3.033195}{Phys. Rev.
  Res. {\bf 3}, 033195}~(2021).

\bibitem{van2021measurement}
Barnaby van Straaten and B{\'a}lint Koczor.
\newblock ``Measurement cost of metric-aware variational quantum algorithms''.
\newblock \href{https://dx.doi.org/10.1103/prxquantum.2.030324}{PRX Quantum
  {\bf 2}, 030324}~(2021).

\bibitem{Grant2019initialization}
Edward Grant, Leonard Wossnig, Mateusz Ostaszewski, and Marcello Benedetti.
\newblock ``An initialization strategy for addressing barren plateaus in
  parametrized quantum circuits''.
\newblock \href{https://dx.doi.org/10.22331/q-2019-12-09-214}{{Quantum} {\bf
  3}, 214}~(2019).

\bibitem{volkoff2021large}
Tyler Volkoff and Patrick~J Coles.
\newblock ``Large gradients via correlation in random parameterized quantum
  circuits''.
\newblock \href{https://dx.doi.org/10.1088/2058-9565/abd891}{Quantum Sci.
  Technol. {\bf 6}, 025008}~(2021).

\bibitem{stokes2020quantum}
James Stokes, Josh Izaac, Nathan Killoran, and Giuseppe Carleo.
\newblock ``Quantum natural gradient''.
\newblock \href{https://dx.doi.org/10.22331/q-2020-05-25-269}{Quantum {\bf 4},
  269}~(2020).

\bibitem{khairy2020learning}
Sami Khairy, Ruslan Shaydulin, Lukasz Cincio, Yuri Alexeev, and Prasanna
  Balaprakash.
\newblock ``Learning to optimize variational quantum circuits to solve
  combinatorial problems''.
\newblock \href{https://dx.doi.org/10.1609/aaai.v34i03.5616}{Proceedings of the
  {AAAI} Conference on Artificial Intelligence {\bf 34}, 2367--2375}~(2020).

\bibitem{gilyen2019optimizing}
Andr{\'{a}}s Gily{\'{e}}n, Srinivasan Arunachalam, and Nathan Wiebe.
\newblock ``Optimizing quantum optimization algorithms via faster quantum
  gradient computation''.
\newblock In Proceedings of the Thirtieth Annual {ACM}-{SIAM} Symposium on
  Discrete Algorithms.
\newblock \href{https://dx.doi.org/10.1137/1.9781611975482.87}{Pages
  1425--1444}.
\newblock Society for Industrial and Applied Mathematics~(2019).

\bibitem{ostaszewski2021reinforcement}
Mateusz Ostaszewski, Lea~M. Trenkwalder, Wojciech Masarczyk, Eleanor Scerri,
  and Vedran Dunjko.
\newblock ``Reinforcement learning for optimization of variational quantum
  circuit architectures''~(2021).
\newblock  \href{http://arxiv.org/abs/2103.16089}{arXiv:2103.16089}.

\bibitem{pirhooshyaran2020quantum}
Mohammad Pirhooshyaran and Tamas Terlaky.
\newblock ``Quantum circuit design search''~(2020).
\newblock  \href{http://arxiv.org/abs/2012.04046}{arXiv:2012.04046}.

\bibitem{fosel2021quantum}
Thomas Fösel, Murphy~Yuezhen Niu, Florian Marquardt, and Li~Li.
\newblock ``Quantum circuit optimization with deep reinforcement
  learning''~(2021).
\newblock  \href{http://arxiv.org/abs/2103.07585}{arXiv:2103.07585}.

\bibitem{rattew2019domain}
Arthur~G. Rattew, Shaohan Hu, Marco Pistoia, Richard Chen, and Steve Wood.
\newblock ``A domain-agnostic, noise-resistant, hardware-efficient evolutionary
  variational quantum eigensolver''~(2019).
\newblock  \href{http://arxiv.org/abs/1910.09694}{arXiv:1910.09694}.

\bibitem{chivilikhin2020mogvqe}
D.~Chivilikhin, A.~Samarin, V.~Ulyantsev, I.~Iorsh, A.~R. Oganov, and
  O.~Kyriienko.
\newblock ``Mog-vqe: Multiobjective genetic variational quantum
  eigensolver''~(2020).
\newblock  \href{http://arxiv.org/abs/2007.04424}{arXiv:2007.04424}.

\bibitem{huang2022robust}
Yuhan Huang, Qingyu Li, Xiaokai Hou, Rebing Wu, Man-Hong Yung, Abolfazl Bayat,
  and Xiaoting Wang.
\newblock ``Robust resource-efficient quantum variational ansatz through an
  evolutionary algorithm''.
\newblock \href{https://dx.doi.org/10.1103/PhysRevA.105.052414}{Phys. Rev. A
  {\bf 105}, 052414}~(2022).

\bibitem{asboth2016schrieffer}
J{\'a}nos~K Asb{\'o}th, L{\'a}szl{\'o} Oroszl{\'a}ny, and Andr{\'a}s P{\'a}lyi.
\newblock ``The su-schrieffer-heeger (ssh) model''.
\newblock In A Short Course on Topological Insulators.
\newblock \href{https://dx.doi.org/10.1007/978-3-319-25607-8}{Pages 1--22}.
\newblock Springer~(2016).

\bibitem{nakanishi2019subspace}
Ken~M Nakanishi, Kosuke Mitarai, and Keisuke Fujii.
\newblock ``Subspace-search variational quantum eigensolver for excited
  states''.
\newblock \href{https://dx.doi.org/10.1103/physrevresearch.1.033062}{Phys. Rev.
  Res. {\bf 1}, 033062}~(2019).

\bibitem{higgott2019variational}
Oscar Higgott, Daochen Wang, and Stephen Brierley.
\newblock ``Variational quantum computation of excited states''.
\newblock \href{https://dx.doi.org/10.22331/q-2019-07-01-156}{Quantum {\bf 3},
  156}~(2019).

\bibitem{mcclean2017hybrid}
Jarrod~R McClean, Mollie~E Kimchi-Schwartz, Jonathan Carter, and Wibe~A
  De~Jong.
\newblock ``Hybrid quantum-classical hierarchy for mitigation of decoherence
  and determination of excited states''.
\newblock \href{https://dx.doi.org/10.1103/physreva.95.042308}{Phys. Rev. A
  {\bf 95}, 042308}~(2017).

\bibitem{santagati2018witnessing}
Raffaele Santagati, Jianwei Wang, Antonio~A Gentile, Stefano Paesani, Nathan
  Wiebe, Jarrod~R McClean, Sam Morley-Short, Peter~J Shadbolt, Damien Bonneau,
  Joshua~W Silverstone, et~al.
\newblock ``Witnessing eigenstates for quantum simulation of hamiltonian
  spectra''.
\newblock \href{https://dx.doi.org/10.1126/sciadv.aap9646}{Sci. Adv. {\bf 4},
  eaap9646}~(2018).

\bibitem{greiner2012quantum}
Walter Greiner and Berndt M{\"u}ller.
\newblock ``Quantum mechanics: symmetries''.
\newblock \href{https://dx.doi.org/10.1007/978-3-662-00902-4}{Springer Science
  \& Business Media}. ~(2012).

\bibitem{mcweeny2002symmetry}
Roy McWeeny.
\newblock ``Symmetry: An introduction to group theory and its applications''.
\newblock Courier Corporation. ~(2002).

\bibitem{sagastizabal2019experimental}
Ramiro Sagastizabal, Xavier Bonet-Monroig, Malay Singh, M~Adriaan Rol,
  CC~Bultink, Xiang Fu, CH~Price, VP~Ostroukh, N~Muthusubramanian, A~Bruno,
  et~al.
\newblock ``Experimental error mitigation via symmetry verification in a
  variational quantum eigensolver''.
\newblock \href{https://dx.doi.org/10.1103/physreva.100.010302}{Phys. Rev. A
  {\bf 100}, 010302}~(2019).

\bibitem{Meyer2022exploiting}
Johannes~Jakob Meyer, Marian Mularski, Elies Gil-Fuster, Antonio~Anna Mele,
  Francesco Arzani, Alissa Wilms, and Jens Eisert.
\newblock ``Exploiting symmetry in variational quantum machine
  learning''~(2022).
\newblock  \href{http://arxiv.org/abs/2205.06217}{arXiv:2205.06217}.

\bibitem{liu2019variational}
Jin-Guo Liu, Yi-Hong Zhang, Yuan Wan, and Lei Wang.
\newblock ``Variational quantum eigensolver with fewer qubits''.
\newblock \href{https://dx.doi.org/10.1103/physrevresearch.1.023025}{Phys. Rev.
  Res. {\bf 1}, 023025}~(2019).

\bibitem{barkoutsos2018quantum}
Panagiotis~Kl Barkoutsos, Jerome~F Gonthier, Igor Sokolov, Nikolaj Moll, Gian
  Salis, Andreas Fuhrer, Marc Ganzhorn, Daniel~J Egger, Matthias Troyer,
  Antonio Mezzacapo, et~al.
\newblock ``Quantum algorithms for electronic structure calculations:
  Particle-hole hamiltonian and optimized wave-function expansions''.
\newblock \href{https://dx.doi.org/10.1103/physreva.98.022322}{Phys. Rev. A
  {\bf 98}, 022322}~(2018).

\bibitem{wang2009efficient}
Hefeng Wang, S~Ashhab, and Franco Nori.
\newblock ``Efficient quantum algorithm for preparing molecular-system-like
  states on a quantum computer''.
\newblock \href{https://dx.doi.org/10.1103/physreva.79.042335}{Phys. Rev. A
  {\bf 79}, 042335}~(2009).

\bibitem{seki2020symmetry}
Kazuhiro Seki, Tomonori Shirakawa, and Seiji Yunoki.
\newblock ``Symmetry-adapted variational quantum eigensolver''.
\newblock \href{https://dx.doi.org/10.1103/physreva.101.052340}{Phys. Rev. A
  {\bf 101}, 052340}~(2020).

\bibitem{Gard2020}
Bryan~T. Gard, Linghua Zhu, George~S. Barron, Nicholas~J. Mayhall, Sophia~E.
  Economou, and Edwin Barnes.
\newblock ``Efficient symmetry-preserving state preparation circuits for the
  variational quantum eigensolver algorithm''.
\newblock \href{https://dx.doi.org/10.1038/s41534-019-0240-1}{Npj Quantum Inf.
  {\bf 6}, 10}~(2020).

\bibitem{barron2021preserving}
George~S Barron, Bryan~T Gard, Orien~J Altman, Nicholas~J Mayhall, Edwin
  Barnes, and Sophia~E Economou.
\newblock ``Preserving symmetries for variational quantum eigensolvers in the
  presence of noise''.
\newblock \href{https://dx.doi.org/10.1103/physrevapplied.16.034003}{Phys. Rev.
  Appl. {\bf 16}, 034003}~(2021).

\bibitem{zhang2021shallow}
Feng Zhang, Niladri Gomes, Noah~F Berthusen, Peter~P Orth, Cai-Zhuang Wang,
  Kai-Ming Ho, and Yong-Xin Yao.
\newblock ``Shallow-circuit variational quantum eigensolver based on
  symmetry-inspired hilbert space partitioning for quantum chemical
  calculations''.
\newblock \href{https://dx.doi.org/10.1103/physrevresearch.3.013039}{Phys. Rev.
  Res. {\bf 3}, 013039}~(2021).

\bibitem{zheng2021speeding}
Han Zheng, Zimu Li, Junyu Liu, Sergii Strelchuk, and Risi Kondor.
\newblock ``Speeding up learning quantum states through group equivariant
  convolutional quantum ansätze''~(2021).
\newblock  \href{http://arxiv.org/abs/2112.07611}{arXiv:2112.07611}.

\bibitem{ryabinkin2018constrained}
Ilya~G Ryabinkin, Scott~N Genin, and Artur~F Izmaylov.
\newblock ``Constrained variational quantum eigensolver: Quantum computer
  search engine in the fock space''.
\newblock \href{https://dx.doi.org/10.1021/acs.jctc.8b00943}{J. Chem. Theory
  Comput. {\bf 15}, 249--255}~(2018).

\bibitem{taube2006new}
Andrew~G Taube and Rodney~J Bartlett.
\newblock ``New perspectives on unitary coupled-cluster theory''.
\newblock \href{https://dx.doi.org/10.1002/qua.21198}{International journal of
  quantum chemistry {\bf 106}, 3393--3401}~(2006).

\bibitem{o2016scalable}
Peter~JJ O’Malley, Ryan Babbush, Ian~D Kivlichan, Jonathan Romero, Jarrod~R
  McClean, Rami Barends, Julian Kelly, Pedram Roushan, Andrew Tranter, Nan
  Ding, et~al.
\newblock ``Scalable quantum simulation of molecular energies''.
\newblock \href{https://dx.doi.org/10.1103/physrevx.6.031007}{Phys. Rev. X {\bf
  6}, 031007}~(2016).

\bibitem{romero2018strategies}
Jonathan Romero, Ryan Babbush, Jarrod~R McClean, Cornelius Hempel, Peter~J
  Love, and Al{\'a}n Aspuru-Guzik.
\newblock ``Strategies for quantum computing molecular energies using the
  unitary coupled cluster ansatz''.
\newblock \href{https://dx.doi.org/10.1088/2058-9565/aad3e4}{Quantum Sci.
  Technol. {\bf 4}, 014008}~(2018).

\bibitem{wecker2015progress}
Dave Wecker, Matthew~B Hastings, and Matthias Troyer.
\newblock ``Progress towards practical quantum variational algorithms''.
\newblock \href{https://dx.doi.org/10.1103/physreva.92.042303}{Phys. Rev. A
  {\bf 92}, 042303}~(2015).

\bibitem{Liu1989}
Dong~C. Liu and Jorge Nocedal.
\newblock ``On the limited memory bfgs method for large scale optimization''.
\newblock \href{https://dx.doi.org/10.1007/BF01589116}{Mathematical Programming
  {\bf 45}, 503--528}~(1989).

\bibitem{mcclean2018barren}
Jarrod~R McClean, Sergio Boixo, Vadim~N Smelyanskiy, Ryan Babbush, and Hartmut
  Neven.
\newblock ``Barren plateaus in quantum neural network training landscapes''.
\newblock \href{https://dx.doi.org/10.1038/s41467-018-07090-4}{Nat. Commun.
  {\bf 9}, 1--6}~(2018).

\bibitem{nakata2017unitary}
Yoshifumi Nakata, Christoph Hirche, Ciara Morgan, and Andreas Winter.
\newblock ``Unitary 2-designs from random x-and z-diagonal unitaries''.
\newblock \href{https://dx.doi.org/10.1063/1.4983266}{J. Math. Phys. {\bf 58},
  052203}~(2017).

\bibitem{Vatan2004}
Farrokh Vatan and Colin Williams.
\newblock ``Optimal quantum circuits for general two-qubit gates''.
\newblock \href{https://dx.doi.org/10.1103/PhysRevA.69.032315}{Phys. Rev. A
  {\bf 69}, 032315}~(2004).

\bibitem{havlivcek2019supervised}
Vojt{\v{e}}ch Havl{\'\i}{\v{c}}ek, Antonio~D C{\'o}rcoles, Kristan Temme,
  Aram~W Harrow, Abhinav Kandala, Jerry~M Chow, and Jay~M Gambetta.
\newblock ``Supervised learning with quantum-enhanced feature spaces''.
\newblock \href{https://dx.doi.org/10.1038/s41586-019-0980-2}{Nature {\bf 567},
  209--212}~(2019).

\bibitem{garcia2013swap}
Juan~Carlos Garcia-Escartin and Pedro Chamorro-Posada.
\newblock ``Swap test and hong-ou-mandel effect are equivalent''.
\newblock \href{https://dx.doi.org/10.1103/physreva.87.052330}{Phys. Rev. A
  {\bf 87}, 052330}~(2013).

\bibitem{cincio2018learning}
Lukasz Cincio, Yi{\u{g}}it Suba{\c{s}}{\i}, Andrew~T Sornborger, and Patrick~J
  Coles.
\newblock ``Learning the quantum algorithm for state overlap''.
\newblock \href{https://dx.doi.org/10.1088/1367-2630/aae94a}{New J. Phys. {\bf
  20}, 113022}~(2018).

\bibitem{kuroiwa2021penalty}
Kohdai Kuroiwa and Yuya~O Nakagawa.
\newblock ``Penalty methods for a variational quantum eigensolver''.
\newblock \href{https://dx.doi.org/10.1103/physrevresearch.3.013197}{Phys. Rev.
  Res. {\bf 3}, 013197}~(2021).

\bibitem{Chufan2022longrange}
Chufan Lyu, Xiaoyu Tang, Junning Li, Xusheng Xu, Man-Hong Yung, and Abolfazl
  Bayat.
\newblock ``Variational quantum simulation of long-range interacting
  systems''~(2022).
\newblock  \href{http://arxiv.org/abs/2203.14281}{arXiv:2203.14281}.

\bibitem{Chufan2022symcode}
Chufan Lyu.
\newblock ``Codes for symmetry enhanced variational quantum spin eigensolver''.
\newblock
  \url{https://gitee.com/mindspore/mindquantum/tree/research/paper_with_code/symmetry_enhanced_variational_quantum_spin_eigensolver}~(2022).

\end{thebibliography}

\onecolumn
\appendix

\section{Ansatz symmetry proof}

Here, we prove the ansatzes shown in Fig.~\ref{fig:ansatz}(c) and Fig.~\ref{fig:ansatz}(d) conserve $S_z$ symmetry and $S_{tot}$ symmetry, respectively. We note that, $\sigma_x^1 \sigma_x^2 + \sigma_y^1 \sigma_y^2 + \sigma_z^1 \sigma_z^2 = 2 \mathcal{P}_{1,2} - I$, where $\mathcal{P}_{1,2}$ is the swap gate, which is defined between any two qubits $m$ and $n$ as $\mathcal{P}_{m,n} \ket{\psi_m \phi_n}=\ket{\phi_m \psi_n}$. Thus, we have
\begin{align}
  \mathcal{N}_{1, 2}(\theta) &= e^{i \theta (\sigma_x^1 \sigma_x^2 + \sigma_y^1 \sigma_y^2 + \sigma_z^1 \sigma_z^2)} \nonumber \\ 
  &= e^{i \theta (2 \mathcal{P}_{1,2} - I)} \nonumber \\
  &= e^{i 2 \theta \mathcal{P}_{1,2}}e^{-i \theta I} \nonumber \\
  &= e^{-i\theta}(\cos{2\theta} I + i\sin{2\theta} \mathcal{P}_{1,2}).
  \label{eq:N_func_1}
\end{align}
Notice that, for any global spin operator $S_\alpha=\frac{1}{2}\sum_i \sigma_{\alpha}^i (\alpha=x,y,z)$, one can easily show $\mathcal{P}_{m,n} S_{\alpha} = S_{\alpha}$ for every choice of $m$ and $n$. This immediately implies that $[\mathcal{P}_{m,n}, S_{\alpha}^2]=0$. Therefore, it is straight forward to show that $[\mathcal{N}_{m,n}(\theta), S_{tot}^2]=0$. Consequently, the ansatz shown in Fig.~\ref{fig:ansatz}(d), which is a combination of several $\mathcal{N}_{m,n}(\theta)$'s, is $S_{tot}$ symmetry conserving.
On the other hand, the phase shift gate $P(\theta)$ commutes with $\sigma_z$, namely $[P(\theta), \sigma_z]=0$, but do not commute with $\sigma_x$ and $\sigma_y$ ($[P(\theta), \sigma_x]\neq 0, [P(\theta), \sigma_y]\neq 0$). Therefore, The ansatz shown in Fig.~\ref{fig:ansatz}(c) is only $S_{z}$ symmetry conserving.

\end{document}